\documentclass{pasa}%

\usepackage{graphicx}
\usepackage[T1]{fontenc}
\usepackage[utf8]{inputenc}
\usepackage{multicol}
\usepackage{multirow}
\usepackage{times,amsmath,amssymb,longtable,breqn}
\usepackage{pdflscape}
\usepackage[varg]{txfonts}
\usepackage{grffile}
\usepackage{color}
\usepackage{xcolor}
\usepackage{ulem}
\usepackage{hyperref}
\usepackage{comment}
\usepackage{ulem}
\usepackage{soul}
\usepackage{tablefootnote}

\usepackage{enumitem}
\setlist[itemize]{leftmargin=5mm}
\setlist[enumerate]{leftmargin=5mm}

\newcommand\Msun{M_\odot}
\newcommand\rs[1]{_\mathrm{#1}}
\newcommand\tch{t\rs{ch}}
\newcommand\Esn{E\rs{sn}}

\newcommand\timplo{t\rs{implo}}

\title[The evolutionary path of PWNe]{From young to old: the evolutionary path of Pulsar Wind Nebulae}

\author[B. Olmi \&  N. Bucciantini]{Barbara Olmi$^{1,2}$\thanks{barbara.olmi@inaf.it} \, and Niccolò Bucciantini$^{2,3,4}$\thanks{niccolo.bucciantini@inaf.it} 
\affil{$^1$INAF -- Osservatorio Astronomico di Palermo, Piazza del Parlamento 1, 90134 Palermo, Italy}%
\affil{$^2$INAF -- Osservatorio Astrofisico di Arcetri, Largo Enrico Fermi 5, 50125 Firenze, Italy}
\affil{$^3$Dipartamento di Fisica e Astronomia, Universit\`a degli Studi di Firenze, Via G. Sansone 1, I-50019 Sesto F. no (Firenze), Italy}
\affil{$^4$INFN - Sezione di Firenze, Via G. Sansone 1, I-50019 Sesto F. no (Firenze), Italy}}

\jid{PASA}
\doi{10.1017/pas.\the\year.xxx}
\jyear{\the\year}

\usepackage{aas_macros}
\usepackage{hyperref} 
\hypersetup{colorlinks,citecolor=blue,linkcolor=blue,urlcolor=blue}

\hypersetup{draft}

\begin{document}

\begin{frontmatter}
\maketitle

\begin{abstract}
Pulsar wind nebulae are fascinating systems, and archetypal sources for high-energy astrophysics in general. 
Due to their vicinity, brightness, to the fact that they shine at multi-wavelengths, and especially to their long-living emission at gamma-rays, modelling their properties is particularly important for the  correct interpretation of the visible Galaxy. A complication in this respect is the variety of properties and morphologies they show at different ages. 
Here we discuss the differences among the evolutionary phases of pulsar wind nebulae, how they have been modeled in the past and what progresses have been recently made. 
We approach the discussion from a phenomenological, theoretical (especially numerical) and observational point of view, with particular attention to the most recent results and open questions about the physics of such intriguing sources.\\
\newline
\newline
Accepted for publication in PASA, 2023 January 30. Received 2022 December 14; in original form 2022 August 16.
\end{abstract}

\end{frontmatter}

\section{INTRODUCTION }
\label{sec:intro}
The violent death of a massive star ($M\gtrsim 8 \Msun$) as a supernova (SN) is thought to leave behind a compact remnant, in many cases in the form of a rotating and magnetized neutron star, known as pulsar (PSR).
Pulsars have outflows, blowing out from the open region of their magnetosphere in the form of  highly relativistic (with Lorentz factor $\Gamma_w \gg 1$), magnetized (with ratio between the Poynting and kinetic energy fluxes $\sigma \gg 1$) and cold plasma winds. 
This wind blows inside the cold debris of the parent star (the ejecta), themselves slowly expanding (with typical speeds $\sim 300-5000$ km s$^{-1}$, \citealt{Chevalier:1976}) in the outer inter-stellar medium (ISM). 
The interaction between the relativistic pulsar wind and these confining ejecta, gives rise to the formation of a wind bubble bounded in the inside by a strong magneto-hydrodynamic (MHD) termination shock  (TS). This  literally  terminates the pulsar wind: at its surface the plasma is heated due to the randomization of the bulk motion, magnetic field is dissipated, and particles are  accelerated \citep{Gaensler:2006, Slane:2017}.
This scenario typically applies to young pulsar wind nebulae but, as we will see in the following, a rather similar one can explain the formation of bow shock nebulae created by the outflow emanating from an (older) pulsar directly interacting with the ISM.
%
\begin{figure}
\centering
	\includegraphics[width=.45\textwidth]{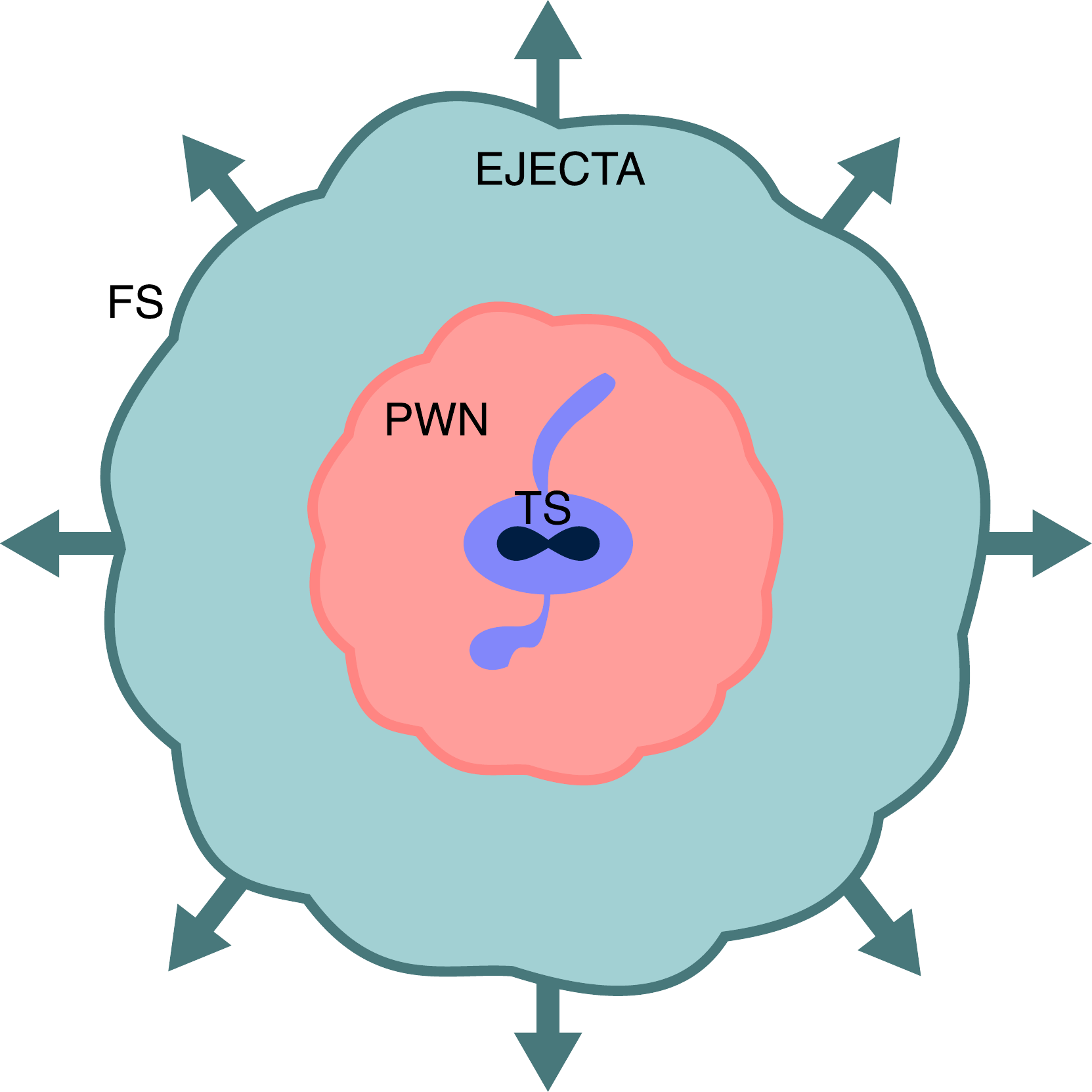}
    \caption{Sketch of the structure of a PWN: the nebula (in light red) is embedded in the expanding SNR ejecta (in aquamarine color), separated from the outer ISM by the SN forward shock. The darker region at the PWN center mimics the pulsar wind termination shock, a region producing no emission and thus visible as an under-luminous area in the inner nebula. The inner structure of the PWN, characterized by a jet-torus morphology visible at X-rays, is drawn in light blue.}
    \label{fig:PWNsketch}
\end{figure}

The pulsar wind nebula (PWN) becomes observable due to the non-thermal emission of particles accelerated at the shock. 
For young systems most of the energy is radiated as synchrotron emission from the relativistic $e^--e^+$ pairs spiralling in the local magnetic field.
Higher energy gamma-rays, in the GeV to TeV ranges, are instead produced by inverse Compton scattering (IC) processes between the same relativistic  particles responsible for synchrotron, and various background photons (either synchrotron photons, photons from the infrared background, photons of the cosmic microwave background, starlight photons). 
The high-energy spectrum might extend up to hundreds of TeV, where Klein-Nishina suppression produces a natural cut-off.
The appearance of old systems, as we will detail in the following, might be very different, both due to lower injection of energy  from the pulsar, the synchrotron cooling and the value of the nebular magnetic field (with IC possibly even exceeding the synchrotron emission).

Despite the success of purely leptonic models in reproducing the spectral energy distribution of many PWNe,  room is left for the possible presence of hadronic emission, produced in the decay of neutral pions following proton-proton collisions. 
The presence of hadrons in the pulsar wind is still matter of discussion \citep{Atoyan:1996,Bednarek:1997,Amato:2003,Bednarek:2003},  and a clear evidence for their existence, or proof of their absence, is missing. 
For example, the recent detection by the LHAASO experiment of photons with energy above $1$ PeV (i.e. $10^{15}$ eV) in coincidence with the position of the Crab nebula \citep{LHAASO_crab:2021}, has been claimed as
a possible indication of high energy protons in the Crab wind. 
Unfortunately, despite this very impressive result, the lack of a robust statistics above 500 TeV makes the overall spectrum still compatible with a fully leptonic scenario (for a more in depth discussion see \citealt{Amato_Olmi:2021}). 

Much of the importance of PWNe is tied to the fact that the Crab nebula, the prototype of this class, is the brightest non-thermal object in the sky over almost its entire electromagnetic spectrum, from radio, to optical, X-rays and beyond. As such, it offers us a unique laboratory where high energy astrophysical processes can be investigated in great details.
On the other hand, the fact that this class includes many other systems, even if fainter and not as well constrained, makes it possible to generalize many of the findings.
From the morphological point of view, PWNe appear as fill-center objects, with possibly a darker region surrounding the pulsar in the very inner nebula that marks the pulsar wind zone (the dark blue region in the sketch of Fig.~\ref{fig:PWNsketch}). Due to its physical properties, the pulsar wind does not produce any sizeable amount of radiation (apart perhaps pulsed emission), and it is then extremely hard to infer information on the pulsar outflow properties before it crosses the TS.
In the Crab nebula, the extension of the emitting region decreases with increasing frequency from radio to X-rays as 
a direct effect of the synchrotron process: the particles life-time against synchrotron losses is in fact inversely proportional to their energy. 
Radio emission then appears in general
more extended (and uniform) than the X-ray one, that is mostly confined to the inner nebula, and highlights the complex structure of the local plasma. 
A very rough comparison of the difference in extension can be seen in the  red sketch of Fig.~\ref{fig:PWNsketch} (light red for the radio nebula, light blue for the X-ray one). 
This however is not the case in all the other systems. In MSH15-5\textit{2} the radio, X-ray and gamma-ray sizes are comparable \citep{Gaensler:2002,2005AA...435L..17A,2016sros.confE..53L}. In G21.5-0.9 the infrared (IR) size is smaller than the X-ray one, that indeed is very similar to the radio extension \citep{Guest:2020,2012A&A...542A..12Z}. 
In the Kookaburra PWN the X-ray size is larger than the radio one \citep{1999ApJ...515..712R,2001ApJ...561L.187R}. 
This just to indicate that extrapolating global trends from the Crab nebula might be misleading and that there is a large diversity in the population.

To date around 110\footnote{For an updated catalog see the SNRcat at: \url{snrcat.physics.umanitoba.ca}, presented in \citet{Safi-Harb:2013}.} sources, observed in different energy bands, have been recognized as -- or candidate to be ($\sim 20$ cases) -- PWNe.
Around 20 of those systems do not have an associated pulsar\footnote{A list of X-ray PWNe with no associated pulsar can be found in \citet{Kargaltsev:2017}}.
Most of them have been detected first at X-rays, while the multi-wavelength identification of a PWN is, in general, not an easy task.
The extended radio  emission might be contaminated by diffuse emission from the surroundings, and from radio data it is very difficult to constrain precisely the morphology of the system, which can help to identify it.
A more important reason of this difficulty is the lack of sensitivity to the required large angular scales (typical of old PWNe) of most radio interferometers which do these observations. Optimal strategies would require the combined use of interferometry and single-dish observations \citep{Kurono:2009}.
X-ray emission is then typically the main discriminant for the identification, especially for young PWNe, given that there are not so many other extended Galactic sources bright in this band. 
Since the beginning of the century, mainly thanks to the very impressive sensitivity of the Chandra telescope for X-ray imaging, our ability to derive information about PWNe morphology has improved significantly.
Unfortunately, X-rays  are the first to fade away as time passes by:  X-ray emission only lasts for a (relatively) small fraction of the PWN life. Middle aged PWNe show in fact very limited (if not any) X-ray emission, making them hardly detectable.
On top of this, as the PWN gets larger, issues with the limited field of view of many of the X-ray instruments begin to play a role.

When getting older, PWNe end up being observable mostly at gamma-rays, where the dominant population of emitting particles is the same producing the long lasting radio emission in the synchrotron band.
Unfortunately, the instruments resolution at gamma-rays is much worse than that at lower energies, and it is then very difficult to infer the nature of a source from its morphology.
Present gamma-ray data almost certainly contain a larger number of unidentified PWNe than of identified ones:  14 are the PWNe firmly identified in the last H.~E.~S.~S. Galactic Plane Survey \citep{HESScoll:2018-GPS}, while around 45 are the unidentified sources. Many of these -- if not almost all -- are actually believed to be unidentified PWNe, 
expected to represent the most numerous class of sources in the very high energy sky \citep{de-Ona-Wilhelmi:2013,Klepser:2013,HESScoll:2018-PWNe}.

In the next future we will see the advent of a new generation of Imaging Atmospheric Telescopes (IACTs) for gamma-ray observations, as the Cherenkov Telescope Array (CTA,  \citealt{2011ExA....32..193A}) or the ASTRI\footnote{ASTRI stands for \textit{Astrofisica con Specchi a Tecnologia replicante Italiana}, for more information visit: \url{www.astri.inaf.it/en}.} Mini-Array \citep{ASTRIMA:2022}. 
If compared with water Cherenkov detectors, as LHAASO or Tibet As-$\gamma$, they will operate in a reduced energy range but with a much better sensitivity and resolution. 
As a consequence, the number of PWNe detected at gamma-rays is expected to increase significantly \citep{Fiori:2022, Remy:2022}, posing the problem of the identification of a very large number of new PWNe, mostly in their middle-age stage.

In this review we will discuss the different evolutionary phases of a PWN, from its early quasi-spherical evolution, to the late phases characterized by a completely different shape and possibly by strong asymmetries, presenting available theoretical models, results form numerical studies, observational hints and open problems.

The review is structured as follows: in Section \ref{sec:evo} we qualitatively introduce the properties of the different phases in which we can roughly divide the evolution of a PWN. Here we also consider PWNe evolving in different environments, discussing their nature as prototype high-energy astrophysical sources.
In the following  Section \ref{sec:theory}, we will briefly inspect theoretical and numerical models of PWNe from an historical point of view; we will highlight results from recent 3D MHD models of young and old PWNe.
In Section \ref{sec:rad_and_acc} we describe the radiation processes at the base of the observed multi-wavelength emission from those sources, and discuss the possible acceleration mechanisms producing particles responsible for the observed emission.
A description of the actual sample of PWNe, from a multi-wavelength point of view,is presented in Section \ref{sec:obs}, where we also review the observational properties of systems at different stages of evolution.
Finally, in Section \ref{sec:open_and_where} we discuss the future prospects both in terms of observations and modelling, with particular focus on the actions needed to answer old and new open questions in pulsar and PWNe physics.

\noindent A brief summary of the main arguments treated in this review and our concluding remarks can be found in Section \ref{sec:conclusion}.

\section{EVOLUTIONARY STAGES OF PWNe}
\label{sec:evo}
%
\begin{figure*}
\centering
	\includegraphics[width=.97\textwidth]{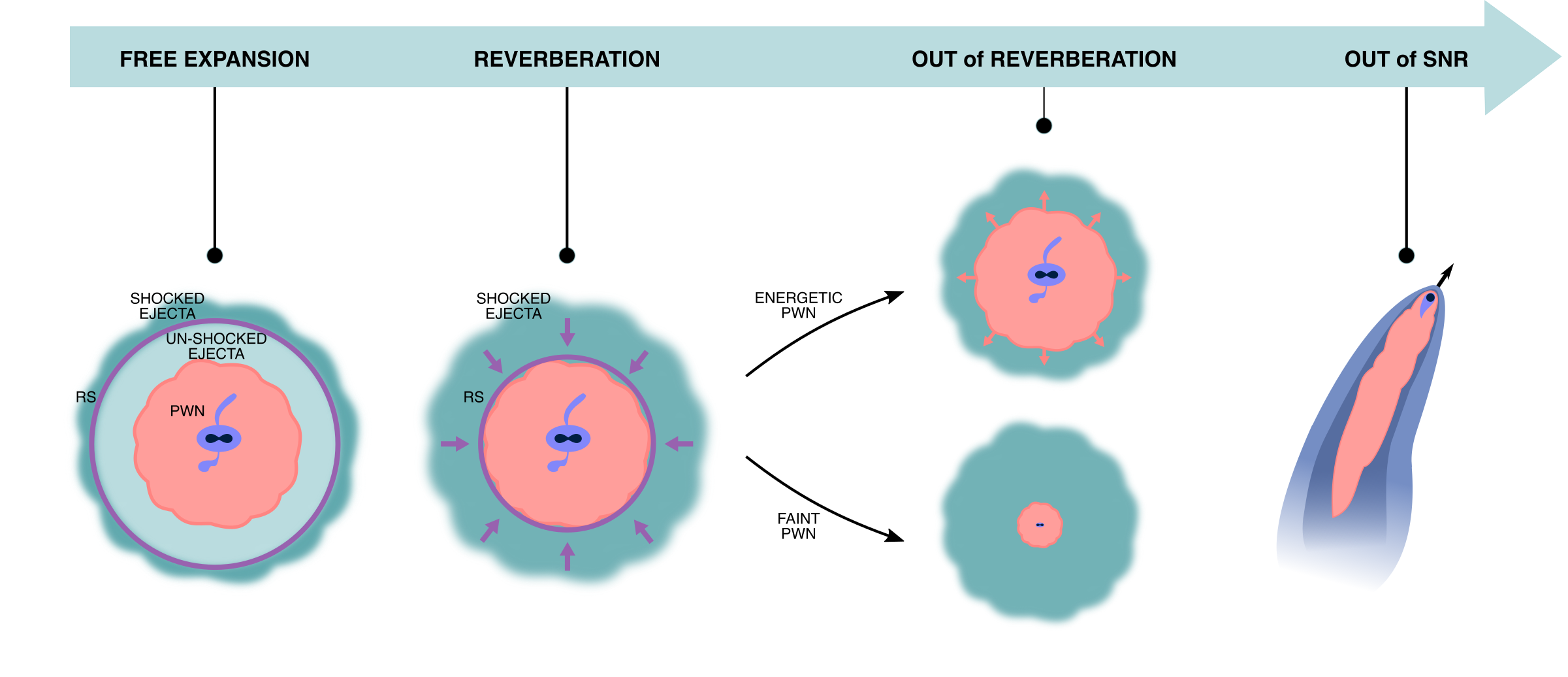}
    \caption{Sketch of the different evolutionary phases of a PWN. From left to right: the PWN expands in the un-shocked ejecta; the RS crushes the PWN and marks the begin of reverberation; the outcome of reverberation depends on the possibility for the PWN to efficiently contrast the compression exerted by SNR; 
    high-velocity enough pulsars eventually exit their parent SNR bubble and interact directly with the ambient medium, producing cometary nebulae (bow shocks). 
    The colors for the PWN ant the ejecta are the same used in Fig.~\ref{fig:PWNsketch}.}
    \label{fig:PWNevo}
\end{figure*}
%
The evolution of a PWN can be ideally divided into four different stages, even if a clear distinction to mark the onset of a new phase from the end of the previous one is hard to make especially for the later stages, which approach a continuum.
A graphical representation of the evolutionary path of a PWN -- and the four phases -- can be seen in Fig.~\ref{fig:PWNevo}. 
The typical duration of each phase depends on both the evolution of the pulsar properties (mainly the spin-down luminosity) and the dynamics of the supernova remnant (SNR) where the PWN expands. 
\subsection{PSR Evolution}
The engine driving the dynamics of  the PWN is the central pulsar.
It loses energy, injecting it in the form of a relativistic plasma, following a typical magnetic dipole spin-down rate (or luminosity, see \citealt{Pacini_Salvati:1973}):
\begin{equation}\label{eq:edot}
    \dot{E} = L = 4 \pi^2 I \frac{\dot{P}}{P^3}\,,
\end{equation}
where $I\simeq 10^{45}$ g cm$^2$ is the pulsar moment of inertia, while $P$ is its rotational period and $\dot{P}$ its time derivative.
For canonical pulsars the period varies typically between  $0.01$~s and $1$~s, while the period derivative is in the range $10^{-15}-10^{-11}$ ss$^{-1}$.
As one might expect, X-ray bright synchrotron nebulae, likely younger, tend to be associated with energetic PSRs ($\dot{E} \gtrsim 10^{36}$~erg~s$^{-1}$). 
Older objects instead tend to be associated with less powerful engines (see e.g. Fig.~\ref{fig:PPdotAll} in the following Sec.~\ref{sec:obs}).

To model the evolution of the pulsar energy injection with time, it is customary to assume that its spin slows down from an initial value $P_0$ according to: $\dot{P} \propto P^{2-n}$, where the coefficient $n$ is called the braking index. This is usually assumed to be in the range $2\leq n\leq3$, with $n=3$ the case of  pure dipole spin-down. However, recently larger values have also been suggested (as in the case of H.~E.~S.~S. J1640-465, reported in \citealt{Archibald:2016}), even if their physical meaning is still not fully understood \citep{Parthasarathy:2020}.
If $n$ remains constant during the pulsar life, the variation of the pulsar energy input is \citep{Pacini_Salvati:1973}:
\begin{equation}\label{eq:Lt}
 L(t) = L_0 \left( 1 + \frac{t}{\tau_0} \right)^{-(n+1)/(n-1)}\,,
\end{equation}
where $\tau_0=P_0/[(n-1)\dot{P_0}]$ is the initial spin-down time of the pulsar.
From this equation it is easy to see  that the pulsar input can be considered as  constant only for $t\ll \tau_0$, while properly accounting for its temporal evolution becomes important for the correct modelling of the nebula beyond $\tau_0$.

\subsection{SNR Evolution}
A very important point that we want to make before discussing the phenomenology of the various evolutionary phases, is that the age of the PWN (or the pulsar) is not a good indicator of the evolutionary stage of the PWN.
A much better indication is instead given by the SNR characteristic time, first introduced by \citet{T&M99}, providing the typical time scale for  the evolution of the SNR:
\begin{equation}\label{eq:tch}
    t_{\rm{ch}} =E_{\rm{sn}}^{-1/2} M_{\rm{ej}}^{5/6} \rho_{\rm{ism}}^{-1/3}\,,
\end{equation}
where $E_{\rm{sn}}$ is the SN explosion energy (typically of order of $10^{51}$ erg),  $M_{\rm{ej}}$ is the mass of the ejecta (in the range $\sim6-20 \Msun$ for SNRs hosting a PWN, see e.g. \citealt{Smartt:2009}) and $\rho_{\rm{ism}}$ the density of the interstellar medium (ISM).

We wish to remark here that, contrary to the widely used \citet{T&M99} model, which predicts large variations, depending on the progenitor structure, in the evolution of the SNR reverse shock (RS, an important factor in setting the various PWN evolutionary stages), and in particular on the time it takes to reach the center of the remnant (an event that we name implosion, happening at time $\timplo$), a much smoother trend has been recently found by \citet{Bandiera:2021}, using fully Lagrangian simulations.
In particular, large variations of the implosion time only appear when drastically chancing the structure of the ejecta core (see e.g. Fig.~4 of that work), while for the largely used case in PWN+SNR models (ejecta with a flat core plus a steep envelope), implosion happens at $\timplo\simeq 2.4 \, \tch$. 
This value clearly represents the maximum time the PWN can remain in free-expansion (see next subsection), and can be used as a sort of theoretical upper limit for the duration of this first phase see Sec.~\ref{sub:rev}.
In reality, for typical PSR energy injection,  the duration of the free-expansion phase is no longer than about half of the implosion time. The implosion time is more representative of the time at which the compression due to the interaction of the PWN with the SNR reaches the maximum.
It should be clear that, depending on the properties of the SNR, this time might be very different for diverse systems, and this is the reason why the age of the PWN (or the pulsar) is not a good indicator of the stage of its evolution.
Indeed, a change by a factor of 2 in the mass of the ejecta leads to a 1.8 factor of difference in the implosion time: if we assume, for example, a fixed SN energy of $E_{\rm{sn}}=10^{51}$ erg, and an ISM with number density  $1$ particle cm$^{-3}$, a PWN in a $6\Msun$ SNR could stay at most for $\sim5000$ yr in the first stage, while a PWN powered by the same pulsar in a $10\Msun$ remnant could stay in free-expansion for $\sim9000$ yr.

\citet{Bandiera:2023} show that the entire evolution of each PWN-SNR system is fully determined -- in the absence of significative radiative losses -- by the two quantities: $[\tau_0/\tch\,,\; L_0\tch/\Esn]$.
These in practice weight the pulsar time and energetics with respect to the SNR ones.

In the following of this section we limit ourselves to a phenomenological discussion of the properties of the various phases. 
%
%

\subsection{Free-expansion}
\label{sub:freeEXP}
In the first phase  the PWN expands, with a mild acceleration, in the cold freely-expanding ejecta of the SNR core, hence its name. The typical PWN expansion speed is of the order of few thousands km s$^{-1}$, much higher than the average velocity the PSR can acquire during the SN event (the kick velocity, ranging  in 100-500 km s$^{-1}$, see e.g. \citealt{FGK:2006}), so that one can safely consider the PSR to be stationary at this stage.
During this phase the evolution of the PWN is independent from that of the SNR shell, since no interaction has been established yet between the two \citep{Reynolds_Chevalier:1984}.
Based on the results by \citet{Jun:1998}, showing that the PWN collects material in a thin-shell at its boundary during its initial expansion ($t\ll \tau_0$), a simplified description of the PWN evolution can be obtained using the following formulas:
\begin{eqnarray}
\frac{d}{dt}\left(4\pi\,P(t)R(t)^4\right)&= & L(t)\,R(t)\,, \label{eq:pwnEVO1}   \\
\frac{d}{dt}\left(M(t)\frac{dR(t)}{dt}\right)\;\;&=&4\pi\,P(t)R(t)^2
+\frac{dM(t)}{dt}\frac{R(t)}{t},\label{eq:pwnEVO2}
\end{eqnarray}
known as the ``thin-shell approximation''.
Here $P(t)$ is the PWN pressure, $M(t)$ the mass of the thin-shell and $R(t)$ the shell radius, that within the thin-shell approximation, can be taken as that of the PWN.

\begin{figure}
\centering
	\includegraphics[width=.23\textwidth]{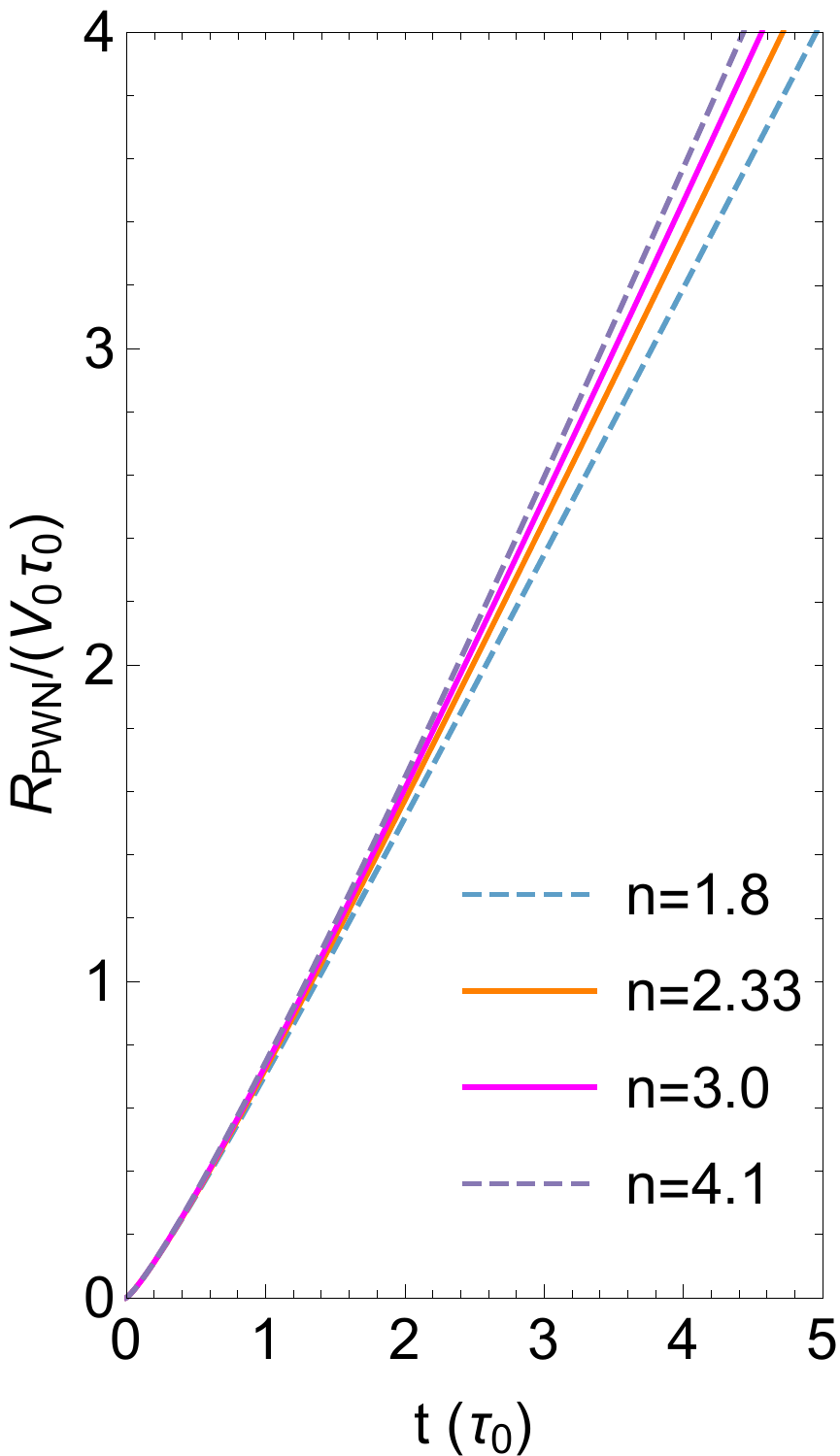}\,
	\hspace{-0.28cm}
	\includegraphics[width=.255\textwidth]{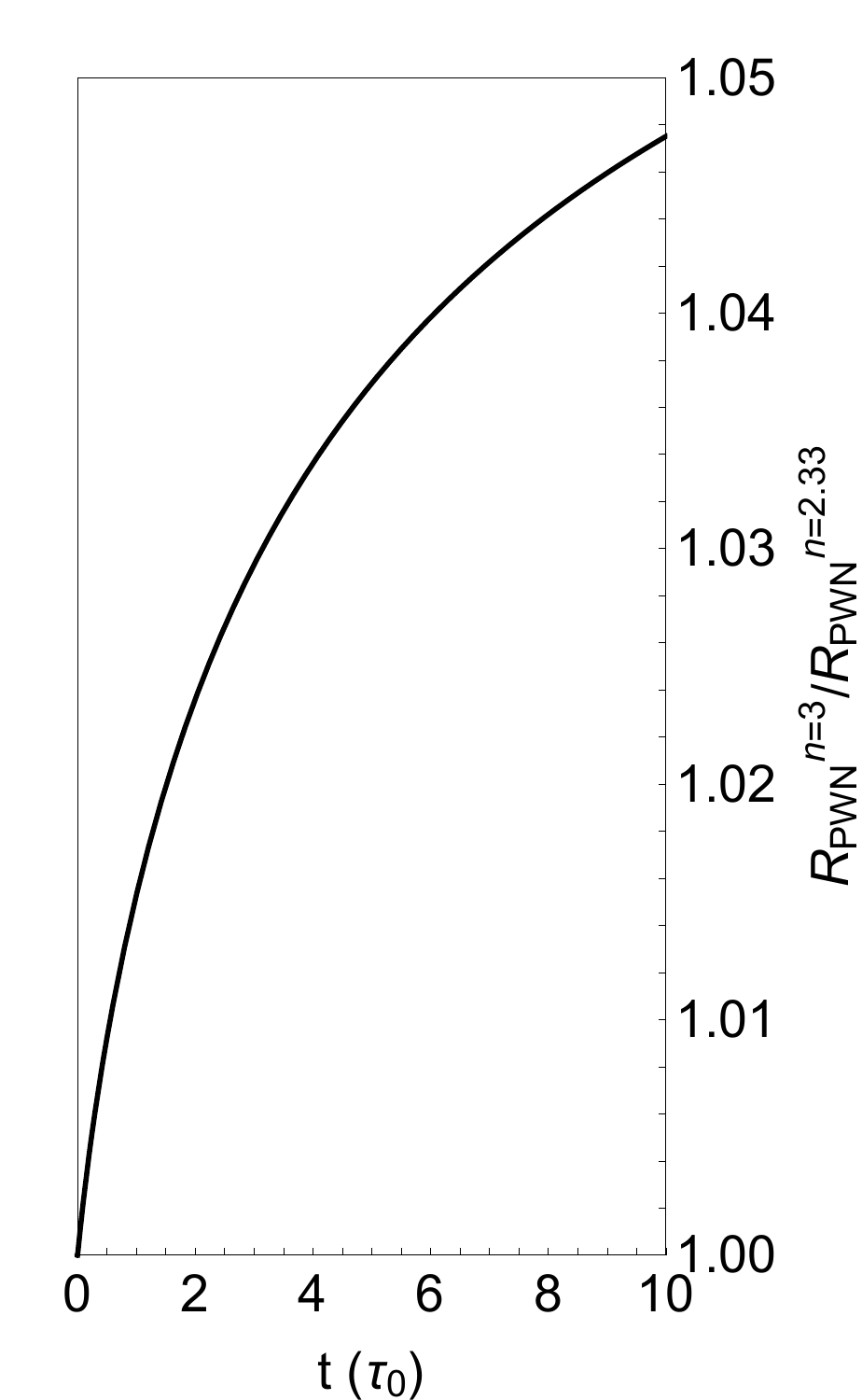}
    \caption{Left panel: time evolution of the PWN radius (in units of $V_0\tau_0$ and $\tau_0$) for four values of the braking index in the range $n=\{1.8;\,4.1\}$. In magenta color we show the pure dipole braking ($n=3$), while in orange the value measured for Crab ($n=2.33$, \citealt{Lyne:2015,Horvath:2019}). Right panel: ratio of the previous two.}
    \label{fig:varying_n}
\end{figure}

The freely-expanding ejecta have a density profile characterized by a flat core surrounded by a steep envelope. It is customary to use  power laws to describe them: the steep envelope as $r^{-\omega}$ (with $\omega > 5$) and the  shallow core as $r^{-\delta}$ (with $\delta<3$, see \citealt{Bandiera:2021} and references therein):
\begin{equation}\label{eq:rhoejprofile}
\rho\rs{ej}(r,t)=
\begin{cases}
A\,(v\rs{t}/r)^\delta/t^{3-\delta}, & \text{if } r < v\rs{t} t\,,	\\
A\,(v\rs{t}/r)^\omega t^{\omega-3}, & \text{if } r \geq v\rs{t} t \,,
\end{cases}
\end{equation}
with the parameters $A$ and $v\rs{t}$ (the expansion velocity of the ejecta core) that  depend on the SN energy and mass of the ejecta as:
\begin{eqnarray}
A &=&\frac{(5-\delta)(\omega-5)}{2 \pi (\omega-\delta)}\frac{\Esn}{v\rs{t}^5}\,,
\\
v\rs{t} &=& \sqrt{\frac{2(5-\delta)(\omega-5)}{(3-\delta)(\omega-3)}\frac{\Esn}{M\rs{ej}}}\,
\,.
\end{eqnarray}
For the simplified case of the flat density profile ($\delta=0$) plus a steep envelope ($\omega=\infty$), actually the most commonly used, the mass of the shell can be expressed as: $M(t)=4\pi R^3(t) A/(3t^3)$.
At early enough times, when the pulsar input can be still considered as constant (namely $t \ll \tau_0$, so that $L(t)\simeq L_0$) an analytical solution for Eq.~\ref{eq:pwnEVO1}-\ref{eq:pwnEVO2} can be found \citep{Bandiera:2023}: 
\begin{equation}\label{Rpwn_vds}
    R\rs{PWN}(t)\Bigg\vert_{t\ll \tau_0}=\left[\frac{(3-\delta)(5-\delta)^3}{(9-2\delta)(11-2\delta)}\frac{L_0}{4\pi}\frac{t^{6-\delta}}{A v_t^\delta}\right]^{1/(5-\delta)}\,.
\end{equation}
This reduces to the well known trend found for the specific case of a flat core ($\delta=0$) by \citet{Reynolds_Chevalier:1984} and later by \citet{van-der-swaluw:2001}: $R\rs{PWN}(t)\propto t^{6/5}$.
On the contrary an analytic solution valid at all times up to reverberation has not been found, while a tentative solution based on the series expansion of the spin-down law was investigated in \citet{Bucciantini:2003}.
For the flat core case, \citet{Bandiera:2023} have recently shown that a simple solution can be determined based on the fitting of numerical results:
\begin{equation}\label{eq:Rpwn_gen}
    R\rs{PWN}(t)\Bigg\vert_{\delta=0} \simeq \mathcal{V}_0\tau_0\frac{\left[ 1+(c_l t/\tau_0 )^{b}\right]^{1/b}}{\left[1+\left(0.82 \,t/\tau_0 \right)^{-a}\right]^{6/(5a)}}\,,
\end{equation}
where the parameters $c_l,\, a$ and  $b$ depend on the value of the braking index $n$, while $\mathcal{V}_0\tau_0\simeq 1.91 (L_0\tau_0/\Esn)^{1/5} (\tau_0/\tch)R_{\mathrm{ch}}$, and $R_{\mathrm{ch}}=M\rs{ej}^{1/3}\rho\rs{ism}^{-1/3}$ is the characteristic radius.
A very interesting result is that the evolution of the PWN in the free-expansion phase is poorly affected by the value of the braking index, at least in the range $1.8 \leq n \leq 4.1$.
This can be seen in Fig.~\ref{fig:varying_n}, comparing the radial evolution computed with Eq.~\ref{eq:Rpwn_gen} for $n=1.8$ ($c_l=0.86310,\, a=0.78492,\, b=0.76490$), $n=2.33$ ($c_l=0.9517,\, a=0.73014,\, b=0.71979$), $n=3$ ($c_l=1.0329,\, a=0.66355,\, b=0.65937$) and $n=4.1$ ($c_l=1.1334,\, a=0.59129,\, b=0.59210$), up to a maximum time of $5-10 \tau_0$, much longer than the duration of this initial phase for typical systems.
%

As we will better see in Sec.~\ref{sec:theory}, being characteristic of young objects, and hence of bright systems, the free-expansion phase has received a lot of attention in the past in terms of modelling, with a variety of approaches  based on different strategies  \citep{Reynolds_Chevalier:1984,van-der-swaluw:2001, Bucciantini:2003, Gelfand:2009,Martin:2012,Fiori:2022, Komissarov:2004,Del-Zanna:2006,Porth:2014,Olmi:2014,Olmi:2016}.

\subsection{Reverberation}
\label{sub:rev}
When the PWN outer boundary hits the SNR RS, the free-expansion phase ends and reverberation starts. 
In reality, due to Rayleigh-Taylor like instabilities (R-T), that develop at the PWN boundary, as well as asymmetries of different nature (e.g. high proper motion of the pulsar, strong ISM density gradients, asymmetries induced from the stellar explosion), this transition is not as sharp as simplified 1D models would suggest, neither is the following reverberation phase. 
This phase and its transition have received much less attention due to several reasons: their extreme complexity, the fact that the SNR evolution and properties become more relevant, the PSR kick velocity begins to play a role.
Moreover, there are few well characterized systems from an observational point of view that can be taken as benchmark to evaluate the accuracy of a model. 
Only recently a detailed study of the transition between free-expansion and reverberation have been presented  \citep{Bandiera:2023}, still limited to a simplified 1D evolution.
The time at which this transition happens (that we name $t\rs{begrev}$) is technically obtained as the intercept between the RS radius and the PWN one. It depends on the relation between the PWN energetics ($L_{0}\tau_0$) and $\Esn$, and can be computed with some accuracy in the limit $\tau_0 \ll \tch$ using the following formula by \citet{Bandiera:2023}, expressed for simplicity in terms of the variable  $\lambda_E=\log_{10}[{(L_0\tau_0)/\Esn}]$:
\begin{equation}\label{eq:tbegrev}
    \!\!\!\!\!t\rs{begrev}(\lambda_E)\simeq 2.4102\frac{1-\exp(-0.1494+1.1606\,\lambda_E)}{1+\exp(1.6831 + 0.6805\,\lambda_E)}\,t\rs{ch}.
\end{equation}
An extension to larger $\tau_0$ is possible substituting to $L_0\tau_0$ the energetic of a massive shell interacting with the SNR in place of the PWN, as discussed in Sec. 5.1 of \citet{Bandiera:2023}.
The time $t\rs{begrev}$ typically ranges between 1 and 2 characteristic times for most of the systems (see Fig. 2 of \citealt{Bandiera:2023}).

When reverberation starts, the PWN begins to interact directly with the shocked ejecta in the SNR. The pressure exerted by the ejecta generally induces a strong deceleration of the  shell accumulated at the PWN boundary \citep{van-der-swaluw:2001} which, in many cases, turns into a compression of the nebula. 
During this compression the internal energy of the PWN, and its magnetic field, increase. Ultimately, depending on the efficiency of radiation losses and magnetic energy dissipation, the total internal pressure rises to become comparable to the external one, and the compression is suddenly reverted into a new expansion. 

One-zone and 1D models predict a number of successive oscillations and expansions before the system reaches a new equilibrium in the Sedov-Taylor phase, and the PWN keeps expanding without bouncing back anymore. This oscillatory behavior is what gives the name  \textit{reverberation to this} phase.
As already pointed out in \citet{Bucciantini:2011}, this extended  oscillatory behavior is an artifact of the  1D approximation.
Higher dimensional models in fact show that during reverberation the PWN boundary is largely subjected to  R-T like instabilities at the boundary \citep{Blondin:2001, Kolb:2017}, producing effective mixing between the PWN material and the outer one. This acts as a sort of viscous term that might halt oscillations already after the first compression.
This is why in one-zone models reverberation is usually stopped artificially after the first bounce \citep{Martin:2012,Torres:2014}, or when the PWN pressure reaches equilibration with  the outer one  \citep{Bucciantini:2011}.
In simplified 1D models, the effect of the reverberation phase on the dynamics of the PWN can be measured in terms of the \textit{compression factor} (CF), namely the ratio between its maximum radius (close to the begin of reverberation) and the radius at the maximum of compression, i.e the minimum one. More generally the parameter of interest is the change in the total volume taken by the relativistic non-thermal plasma of the PWN, since it translates in a variation of the nebular magnetic field ($B\propto R^{-2}$). 
It is in fact this quantity that directly impacts the chance an old PWN might become visible again in X-rays.

Roughly speaking, the effects of the reverberation can be divided into two extreme cases: (i) the PWN is sufficiently energetic, then the compression is relatively small, or even not appreciable, and the PWN continues its expansion almost undisturbed; (ii) the PWN is weak and then overwhelmed by the SNR pressure, contracting down to very small radii (volumes).
Low energetic systems are the most critical in terms of modelling, since they might undergo violent compression with important modifications of their multi-wavelength spectral properties.
In fact compression enhances the magnetic field and energises particles, leading to an increase in radiative losses. For very high compression efficiency  ($CF\gtrsim 1000$), the PWN can enter a fast cooling regime, where a large fraction of the particle energy is lost, and the particle energy distribution is strongly modified \citep{Torres:2019}.
\citet{Bandiera:2023} show that these extreme behaviour is not expected to be common within the present PWNe population, with only a relatively limited number of objects having possible CF larger than few hundreds.
On the contrary, the vast majority of the systems undergo a rather small compression, with CF ranging from a few to tens.
However \citet{Bandiera:2023} do not consider the effect of radiative losses and thus, this estimate might change depending on the relevance of losses in the first evolutionary phase.

\subsection{Post-Reverberation}
\label{sub:postrev}
The reverberation phase, together with the pulsar proper motion, is what shapes the PWN in the later evolutionary phases. Gradients in the ambient medium density are known to impact the evolution of the SNR reverse shock \citep{2008A&A...478...17F, Kolb:2017} and, in general, one expects 
that as time passes the level of asymmetries in the system grows. 
This implies that simplified one-zone or 1D models became progressively less accurate in the description of these systems, and attempts to include these extra effects \citep{Gelfand:2009} less and less reliable. 
On top of this, the mixing due to R-T instability can be so strong as to disrupt completely the PWN as a coherent object.
It is then clear that a PWN in this stage would likely be very far from being spherically symmetric, rather having a complex distorted shape.
As a consequence, modelling this phase can be very demanding: the need to properly account for asymmetries requires the use of multi-D models. Being the shape and properties of the evolved PWN strongly dependent on its previous history, one has to follow its entire evolution, which implies a large dynamical range in term of temporal and spatial scales. Moreover the morphology of the system, especially in the presence of mixing and instabilities, is very dependent on the model dimensionality and a comprehensive description can only be done in 3D.
A very beautiful example of a 3D model of a largely asymmetric evolved system can be found in  \citet{Kolb:2017}.

Unfortunately there are not that many systems, with a detailed and robust characterization, that can be used to benchmark our theoretical models for this evolutionary phase. Some of them like G327.1-1.1 \citep{Temim:2015,2016ApJ...820..100M}, IC443 \citep{Swartz:2015}, W44 \citep{Frail:1996,Petre:2002}, have been the targets of previous analyses \citep{van-der-swaluw:2004,Bucciantini:2011}, but they are too few to provide a significative sample.

From the one-zone description, one expects that when reverberation ends, with the oscillatory behaviour almost completely dissipated, the PWN relaxes to a steady subsonic expansion \citep{van-der-swaluw:2001}. This was found to happen on a long time-scale of $\sim20\tch$.
This time is possibly even longer, since the relaxation of the contact discontinuity of the SNR has been recently show to happens on times of $\sim 35\tch$, while between $20\tch$ and $30\tch$ small oscillations are still present in the SNR radius \citep{Bandiera:2021}.

\subsection{Late Phase - bow shocks}
\label{sub:bspwne}
Ultimately  PWNe are likely to end their life as bow shock nebulae (BSPWN), due to the fact that a large fraction of pulsars is born with high kick velocity (100-500 km s$^{-1}$, \citealt{Cordes:1998,Arzoumanian:2002,FGK:2006,Sartore:2010,Verbunt:2017}), and as such it is bound to emerge out of the parent SNR before the pulsar spin-down luminosity becomes so weak as to make particle acceleration and non-thermal synchrotron emission negligible.
Considering the typical SNR decelerated expansion of the Sedov-Taylor phase, one can easily estimate in a few tens of thousand of years the time the pulsar takes to emerge the SNR bubble, to be compared with the longer typical age of pulsars in the Galaxy ($\sim 10^6$ yr).
From that moment on, the pulsar interacts directly with the ambient medium. 
Given the high speed of the pulsar, larger than the typical sound speed of the surrounding medium, its motion generally becomes supersonic already inside the SNR \citep{Gaensler:2006}.
IC443 and W44 are exemplary cases of BSPWNe still confined within their parent SNR \citep{Swartz:2015,Frail:1996,van-der-swaluw:2004}.
The supersonic motion induces the formation of a bow shock around the pulsar and its nebula, reshaping drastically the PWN.
Now it does not appear as a bubble anymore but, on the contrary, it assumes an elongated cometary-like shape.

In BSPWNe the pulsar is located at the bright head of a very elongated tail, extending in the direction opposite to the pulsar motion. 
The new morphology of the system depends on the balance of the ram pressure of  the pulsar wind with respect to that of the incoming (in the frame of the PSR) ambient medium. 
This balance actually determined the thickness of the PWN at the head front, representing the characteristic dimension of a BSPWN, known as \textit{stand off distance}:
\begin{equation}
\label{eq:standoff}
    d_0=\sqrt{\frac{L}{4\pi c \,\rho\, v^2\rs{psr}}}\,,
\end{equation}
where $\rho$ is the density of the ambient medium.
Gradients in the ambient medium reflect into asymmetries in the bow shock shape \citep{Vigelius:2007}.
On the contrary, even larger asymmetries in the pulsar wind energy distribution tend to have minor effects on the overall bow-shock shape \citep{Olmi_Bucciantini_2019_1}, remaining mostly concentrated in the head, that is typically not resolved even in the deepest Chandra's observations.

The PSR wind, shocked in the head, is now diverted along the tail, where the plasma becomes less magnetized and more turbulent with increasing distance from the pulsar \citep{Wilkin:1996, Bucciantini_bowsI:2001,Bucciantini_BowsII_2002}. 
How turbulent and magnetized the tail is depends not only on the PSR wind magnetization, but also on the level of asymmetry in its energy distribution, on the inclination of the PSR spin axis with respect to the direction of the kick velocity, and on the relative inclination and strength of the ambient medium magnetic field \citep{Olmi_Bucciantini_2019_1}.

\subsection{Other environments}
\label{sub:other}

PWNe represent proto-typical systems where one can witness the interaction of a relativistic magnetized outflow with a confining surrounding environment. As such they form a paradigm for a variety of different classes of high-energy astrophysical objects. Moreover, PWNe can be found more or less anywhere a NS is active, even if the activity is just a transient one. Indeed, from an historical point of view, they have often been ``the first ones'' where high-energy astrophysical processes have been discovered/studied/understood, and much of the theory developed for their study has found application elsewhere. Here we briefly review some of these other environments.
  
\begin{itemize}
\item PWNe, from the point of view of fluid dynamics, acceleration properties, and emission mechanisms,  are representative of the very broad class of relativistic wind bubbles. The physics that have been thoroughly investigated to explain the observed presence of non-thermal particles or the properties of relativistic outflows at large distances from their engine \citep{1992SvAL...18..356L,1993PhDT........71L,2008IJMPD..17.1669B,2021ApJ...906..105C,2019ApJ...873..120Z,2015SSRv..191..545K,Sironi:2015} have proven to be quite general and with a much larger applicability, in terms of astrophysical systems ranging from gamma ray bursts \citep[GRBs, ][]{1994MNRAS.270..480T} to active galactic nuclei jets and radio lobes \citep{Uz2016,2018MNRAS.477.2849U}.\\

\item Bright persistent X-ray emitting PWNe require a steady  engine, providing the necessary particle and energy injection. It was held for a long time that only canonical PSRs, with their active magnetospheres, capable of supporting pair creation cascades with high multiplicity \citep{2019ApJ...871...12T} could be surrounded by such nebulae. However, recently has emerged a more dynamical picture, where transient PWNe can form and shine, even around neutron stars with inefficient pair creation, as in the case of magnetars and/or rotating radio transients \citep[RRATs, ][]{2013MNRAS.429.2493C,2009ApJ...707L.148V,2016ApJ...827..135T,2017ApJ...835...54T,2017MNRAS.464.4895G,2017ApJ...850L..18B,2018IAUS..337..255T,2018ApJ...868L...4M}. These transient nebulae,  typically associated with bursting/active phases, are still electromagnetic powered, and shine in synchrotron, even if it is still unclear if the neutron star activity leads to freshly injected particles, or simply re-energizes particles already present and injected at earlier times.
Even the driving energy reservoir is not well constrained in general, though magnetic energy has been suggested for the case of magnetars.
Moreover, violent events, as giant flares, can produce transient nebulae, whose evolution resembles, even if at a faster pace, that of regular PWNe \citep{2005Natur.434.1104G,Gelfand:2005}.\\

\item PWNe can also form in binary systems, where the pulsar wind is confined by the ram pressure of the companion star.
In case of massive stars, with strong equatorial flows, this interaction leads to the formation of a bow-shock, whose dynamics, and by consequence emission, can be both highly variable due to the possible large orbital eccentricity, and to the fact that it is also strongly affected by centrifugal and Coriolis forces.
As in normal bow-shocks, particles can be accelerated to high energy, but unlike regular systems, the orbital dynamics can lead to a substantial mixing with the stellar material, a high level of turbulence, and the development of multiple shocks, with distributed acceleration.  Moreover the presence of a bright source makes IC a relatively strong cooling process, to the point that it is unclear if the observed X-ray emission is synchrotron or IC \citep{2007astro.ph..1144N,2014ApJ...784..124K,2013MNRAS.430.2951B,2010MNRAS.403.1873Z,2015A&A...574A..77P,2012A&A...544A..59B}. 
These systems offer us a unique opportunity to investigate the PSR wind and its acceleration properties, at distances much smaller than in regular PWNe, and can have important consequences on a large set of dissipative wind models \citep{2003ApJ...591..366K,2006AdSpR..37.1970K,2010ApJ...725L.234L,2012ApJ...755...76T,2017A&A...607A.134C,2020A&A...642A.204C,2021A&A...646A..91H,2021A&A...649A..71H}. \\

 \item Proto-magnetar wind nebulae, and in general relativistic outflows from millisecond rotating newly born magnetars, have been invoked to explain both long  \citep{1992:usov,2007RMxAC..27...80T,2007MNRAS.380.1541B,2008MNRAS.383L..25B,2011MNRAS.413.2031M} and short GRBs \citep{2012MNRAS.419.1537B,2017AcASn..58...29W,2014MNRAS.439.3916M,2015ApJ...802...95R}, mostly because the detection of the so called ``late activity'' \citep{2014MNRAS.438..240G,2013MNRAS.430.1061R} has made almost mandatory to assume long lived engines, hardly compatible with the timescale for disk accretion in a stellar mass black hole. 
 The key idea is that the relativistic outflow that we think is at the origin of the prompt gamma-ray emission, is nothing else than the magneto-centrifugally driven neutrino-wind, coming from the cooling proto-NS. The dynamics of such wind reaches rapidly the force-free limits, as the baryon loading rapidly drops in time \citep{Metzger_Giannios+11a}, and its interaction with the surrounding layers of the progenitor star or the ejecta of the binary merger (for SGRBs), leads to the formation  of a hot relativistic wind bubble, that is both a reservoir of energy, to be released at later time, as well as a reservoir of high energy photons that can lead to the appearance of a so called kilo-nova \citep{2015ApJ...798L..36C,2016ApJ...819...14S,2016ApJ...819...15S}. 
 Lower energy systems have been considered also to justify a possible continuum of explosion phenomenology down to super-luminous and broad line Ib/c supernovae  \citep{2012MNRAS.426L..76D,2017ApJ...835..177M,2017ApJ...851...95S,2018ApJ...864L..36M,2019RAA....19...63W,2021ApJ...917...77V,2021MNRAS.508.5390S,2019A&A...621A.141D,2018MNRAS.475.2659M}.\\

\end{itemize}
\section{modelling PWNe}
\label{sec:theory}
The pulsar wind of an oblique rotator has been shown to have an extremely complex structure named \textit{striped wind}, with a magnetic field ($B$) of alternate polarities separated by a current sheet \citep{Bogovalov:1999}. 
A very important parameter of the pulsar wind is the wind magnetization $\sigma$, representing the ratio between the Poynting flux and the particle kinetic energy flux in the wind:
\begin{equation}\label{eq:sigma}
    \sigma= \frac{B^2}{4\pi \rho  c^2 \Gamma^2_w}\,,
\end{equation}
with $\rho$ the comoving density of the plasma and $\Gamma_w$ the Lorentz factor of the wind.
In recent years our understanding of the properties of the pulsar wind has been increased thanks to very refined numerical models, going from force-free only \citep[][and subsequent]{Spitkovsky:2006} to more complex physics, including pair creation \citep{Philippov:2014}.

In this section we review the many different approaches that have been used to model PWNe, discussing merits and  pitfalls of each  approach, as well as their validity with respect to the different evolutionary phases.
A complete description of a PWN, to be compared with observations,  requires the treatment of two different aspects: (1) the dynamical evolution of the system; (2) the spectral evolution of the particles responsible for the emission.
At present none of the models is able to account for both at a level of accuracy to enable a direct and reliable comparison with data.
Multi-dimensional numerical simulations have reached in the last years excellent results in the description of the dynamics, but on the other hand they lack in terms of spectral modelling, with radiation  generally evaluated on top of the dynamical results at the end of the computation, with simplified treatment of the radiation losses and of the particle energy evolution \citep{Volpi:2008, Porth:2014, Olmi:2016}.
On the other hand, accurate radiative models only exist for very simplified 1D descriptions of the dynamics, thus the spectral evolution is pinned  to a rough model of the physical properties of the system \citep{Gelfand:2009,Bucciantini:2011,Torres:2014}.

\subsection{One-zone models}
\label{sub:0Dmod}
In one-zone (or equivalently 0+1) models, the PWN is described as a uniform bubble in interaction with the surrounding SNR, being  subjected to adiabatic, radiation and possibly particles losses. Energy and magnetic field are injected into the bubble with constant relative efficiencies by the PSR, following its spin-down luminosity (Eq.~\ref{eq:Lt}). The magnetic field energy within the bubble can be evolved either assuming magnetic flux conservation \citep{Pacini_Salvati:1973,Gelfand:2009,Torres:2014} or a constant ratio between magnetic and plasma pressures \citep{Bucciantini:2011}. 
In the early free-expansion phase, if radiation losses are negligible, the two approaches are equivalent (however they might have different implications in terms of wind magnetization), but once radiation losses become important they  might lead to sizeable different results \citep{Bucciantini:2011}.

One-zone models are all based on the thin-shell approximation: the shell is evolved considering mass and momentum conservation (Eqs.~\ref{eq:pwnEVO1}-\ref{eq:pwnEVO2} in Sec-~\ref{sub:freeEXP}) and its radius is equated with that of the PWN. 
To the time evolution of the radius, one then couples a description of the evolution of the internal particles distribution function subject to injection and losses, from which one finally derives the observed PWN multi-wavelength spectrum.

Spectral evolution models can be traced back to the original work by \citet{Pacini_Salvati:1973} that, assuming an approximated description of the temporal evolution of the PWN, limited to its initial free-expanding phase, derived the evolution of the spectral energy distribution function  showing its relation with the injected particle distribution. 
This work was later extended with a model also describing the interaction with the SNR, but still limited to the interaction with the unshocked ejecta \citep{Reynolds_Chevalier:1984}, and considering possible variations to the particles injection \citep{Atoyan:1999}.

Despite their evident over-simplification, one-zone models have been -- and still are -- widely used.
Results from one-zone models have in fact proved to give a good description of the global properties of young PWNe, allowing one to model the evolution of these systems and their spectral properties, and to sample a large parameter space, in ways that are not possible with current multi-dimensional models. 
They can be easy implemented and are not much demanding in terms of numerical resources \citep{Gelfand:2009, Tanaka:2010,  Bucciantini:2011, Martin:2012, Tanaka:2013, Torres:2014,Gelfand:2017, Fiori:2020}. 
These characteristics make them very much appealing for large populations studies (e.g. \citealt{Torres:2014, Fiori:2022}), but on the other hand they must be used with care to follow the evolution beyond free-expansion.
In \citet{Bandiera:2020,Bandiera:2023} it has been in fact shown that one-zone models predict excessive compressions of the nebula during reverberation, possibly leading to the burn-off of a huge amount of the emitting particles, changing dramatically the spectral evolution.
\citet{Torres:2019} show that in cases of extreme compressions ($CF\gtrsim1000$) a super-efficient phase can appear, when the PWN luminosity  is so enhanced to make it visible again at X-rays at later times.
One-zone models tend to overestimate the PWN compression mainly because the simplified description of the pressure in the SNR, in general assumed to be equal to the central pressure from the Sedov solution, or the pressure at the SNR FS scaled with some arbitrary factor (for a complete description see \citealt{Bandiera:2023}). 
This kind of approximation indeed introduces large errors when the PWN starts to interact with the SNR. 
Recently \citet{Bandiera:2023} have in particular shown that the pressure in the SNR is very far from the Sedov solution during the entire first compression, and that in the aforementioned assumptions lead to a consistent overestimation of the outer pressure. This means that the number of systems undergoing a super-efficient phase, might be much smaller than expected, if a correct description of the SNR pressure is considered during reverberation.

A preliminary attempt to extend one-zone models beyond free-expansion has been recently made by \citet{Fiori:2022},  where the SNR pressure is shaped on the results of 1D hydrodynamic simulations of the PWN-SNR interaction. 
Conversely to \citet{Bandiera:2023}, in this case there was no specific focus in reproducing correctly the dynamical effects of the interaction between the PWN and the surrounding SNR.

The extension of one-zone models is generally limited to the first series of compressions and re-expansions of the PWN, while in few cases they have been extended to longer times halting by hands the oscillations to match the SNR pressure from the Sedov solution \citep{Torres:2019}.

\subsection{1D models}
\label{sub:1Dmod}
In 1D models the evolution and properties  of the PWN and the SNR are given as a function of the radial coordinate.
The first 1D model of a PWN was put forward by \citet{Rees_Gunn:1974}  to describe, as many others later on, the Crab nebula. 
Despite the very simplified description of the system, that model was able to predict a number of features and to give them a physical interpretation:
the appearance of the ultra-relativistic un-shocked wind as an under-luminous area in the inner nebula, later observed by \citet{Weisskopf:2000} at X-rays; the nebula shrinkage with increasing frequency as sign of the synchrotron emission and central particle injection; the nebular magnetic field close to the equipartition value; the Lorentz factor of the wind, its magnetization and the injection rate of particles from the synchrotron luminosity. 
This preliminary model was then elaborated and extended by \citet{K&C:1984a, K&C:1984b}, that provide estimates of many characteristic quantities of the Crab nebula based on the full solution of the relativistic MHD equations within the PWN itself.
Few years later, \citet{Emmering_Chevalier:1987} found a time dependent analytic solution of the same problem.

Many other 1D models have been then developed in later years, all based on the numerical implementation of the equations describing the PWN-SNR interaction and evolution, both in the classic and relativistic regimes and in the hydrodynamic (HD) or MHD frameworks, extending to a longer evolution than the free-expansion phase \citep{van-der-swaluw:2001,van-der-swaluw:2004,Bucciantini:2003,deJager_G21:2008, Bandiera:2023}.

One of the longer standing problems in PWNe physics, the so called \textit{sigma-paradox} \citep{1998MmSAI..69.1009M}, arose as consequence of these 1D models, and puzzled the community for more than thirty years: in order to explain the existence of the TS, the average magnetization $\sigma$ in the nebula must be quite low: $\sigma\simeq \rm{few}\times 10^{-3}$ \citep{K&C:1984b}.
But theoretical models of pulsar magnetospheres predict a much larger magnetization at the light cylinder of the star: $\sigma\sim 10^{4}$ \citep{Kirk:2009,Arons:2012}.
To make compatible these two opposite predictions, a huge amount of magnetic dissipation must be considered along the path separating the light cylinder and the pulsar wind termination shock, converting the pulsar wind from a Poynting dominated outflow to a particle dominated one.
Magnetic dissipation is actually expected to occur in the current sheet of the alternating pulsar wind  \citep{Lyubarsky:2003, Sironi:2011}, but $\sim 7$ orders of magnitude are simply too many to be accounted for with any known process.
We want to remark here that the magnetization at the wind termination shock not only affects the PWN dynamics, but it is also an important parameter that controls the efficiency of the  various acceleration mechanisms \citep{Amato:2014}.
As we will discuss in the following paragraphs, going multi-D is the way to reduce -- if not solve -- this long standing problem, since the augmented dimensionality allows increasing the amount of magnetic dissipation.

Results of one-zone and 1D models are in excellent agreement in the free-expansion phase, while the accordance disappears as soon as the PWN enters reverberation.
As for one-zone models, the reliability of 1D models is in any case limited to at most the beginning of reverberation. Unlike one-zone, however, 1D Lagrangian models \citep{Bandiera:2023} can be extended  to the following evolution (the first compression and the subsequent sequence of multiple compressions and re-expansions). Recall that, in any case, these are just an artifact of the reduced dimensionality.
1D simulations show a sort of damped oscillations, with compressions generally becoming less severe as time passes by. 
These behaviour is not found in a multi-D description of the PWN/SNR system, where the mixing produced by boundary instabilities helps in mitigating oscillations. However, one should be careful not to confuse the observed angular size/dimension of the PWN with its effective volume (the one that matters for compression), once mixing becomes important.
%
\subsection{2D models}
\label{sub:2Dmod}
Overcoming the limitations of 1D models became urgent with the first detailed images of the inner Crab nebula at optical (Hubble Space Telescope, \citealt{Hester:1995}) and  X-rays (Chandra, \citealt{Weisskopf:2000}).
The Crab nebula was in fact the first source where a beautiful bright jet-torus structure, with an equatorial torus and two opposite polar jets, was identified. 
Later on other systems showed up the same properties \citep{Gaensler:2002,Lu:2002,Romani_Ng:2003,Camilo:2004,Slane:2004,Romani_Ng:2005}, now believed to be a common feature of young PWNe. 
It is clear that such a complex structure cannot be reproduced with the simplified geometry of 1D models, neither by slighting modifying the 1D approach.
Indeed the first 2D analytic model of the Crab nebula by \citet{Begelman_Li:1992}, considering the toroidal structure of the magnetic field, was only capable to explain its observed prolate shape. 

A critical point for the theoretical description of the inner nebula was the appearance of the polar jets so close to the pulsar.
They in fact seem to form in the un-shocked relativistic wind, where magnetic collimation is known to be poorly efficient.
Theoretical and numerical models of the relativistic  wind emanating from the pulsar (\citealt{Begelman_Li:1992,Contopoulos:1999, Bogovalov:1999}, and later \citealt{Komissarov:2006, Spitkovsky:2006,Timokhin:2006}) already predicted a non-uniform distribution of the wind energy flux at the termination shock, with most of the energy concentrated in the equatorial plane (the so called \textit{split-monopole} models), but there was no evidence for a collimation of part of the flux in the polar regions.
The solution to the jet formation was found only few years later: modelling the dynamics of the plasma with an anisotropic distribution of the energy flux in the wind, the terminations shock was found to become oblate, with larger extension in the equatorial region than in the polar one \citep{Lyubarsky:2002,Bogovalov:2002a, Bogovalov:2002b, Khangoulian:2003}.
Polar jets can then form due to magnetic hoop stresses in the post-shock plasma, immediately beyond the polar front of the shock, and appear closer to the pulsar than the torus thanks to the oblate morphology of the TS itself.

This theoretical prediction was verified later on with the use of relativistic 2D MHD numerical simulations \citep{Del-Zanna:2004, Komissarov:2004,Bogovalov:2005,Del-Zanna:2006}.
Moreover, 2D models show that the formation of jets is only possible starting from a minimum wind magnetization in the pulsar wind of $\sigma\sim 10^{-2}$, one order of magnitude larger than that originally set by 1D models.
The increase of dimensionality then appears as the first possible way to alleviate the sigma-paradox.

Thanks to its luminosity and vicinity, the largest part of 2D models were made to investigate the properties of the Crab nebula that was -- and still is -- the perfect source to look at for a detailed comparison.
They were extremely successful at reproducing many of its features down to very fine details, especially in the inner nebula: maps of the synchrotron emission \citep{Del-Zanna:2006, Olmi:2014}; polarization properties \citep{Bucciantini:2005}; the complete spectrum, from radio to gamma-rays \citep{Volpi:2008}; the variability at small scales in the inner nebula \citep{Camus:2009,Olmi:2015,Lyutikov:2016}.
In particular the multi-wavelength appearance of variable arc-like bright structures (the so called \textit{wisps}), forming close to the TS and moving outwards at a consistent fraction  of $c$, was shown to be a perfect tool to trace the properties of the underlying plasma and to constrain the location and mechanism of particle acceleration \citep{Olmi:2015}.

On the contrary, results of 2D models at larger scales are limited by the imposed axisymmetry, with only the global  properties of the nebula, such as the extension and the  shrinkage with the increasing energy, reproduced reasonably well.
Axisymmetry is in fact the major limitation of 2D models: it induces the artificial accumulation of magnetic loops along the polar axis of the nebula, producing enhanced compression of the magnetic field.
This imposes an upper limit to the wind magnetization of $\sigma \lesssim 0.1$, otherwise the polar compression of the field becomes so strong that the collimated magnetic flux punches the nebular shell at the polar boundaries.
The consequence of this non-redistribution of magnetic field is that polar regions have an (excessively) intense magnetic field, while the rest of the nebula is under-magnetized, with an average magnetic field, for the Crab, well below the expected value of $\sim150\,\mu$G. 
To reproduce the PWN spectrum then one is forced on one side to inject an excessive number of particles, on the other to steepen by hand the injection spectrum of the X-ray emitting component to alleviate the average lower energy losses, and match the synchrotron emission.

The problem introduced by the wrong geometry of the magnetic field remains very evident if looking at the morphology of the nebula at large scales: the radio emission, expected to be rather uniform, indeed resembles the shape of the magnetic field (it is concentrated around the polar axis); the IC emission is overestimated, approximately by the same factor by which the averaged magnetic field is underestimated \citep{Olmi:2014}.

Despite these limitations, 2D models remain appealing in many cases: their results are in good agreement with more complex 3D models if limited to the inner nebula and small scales; they allow for a longer evolution of the PWN and for a larger investigation in terms of number of sources and physical parameters with respect to their equivalent in 3D, due to the lower numerical cost.
A possibility to use 2D models to infer the large scale properties of PWNe has been investigated in \citet{Olmi_Torres:2020}, where the HD scheme has been preferred to the MHD one to avoid the aforementioned problems with the field geometry, affecting the large scale emission. In that work the magnetic field is excluded from the dynamics but traced numerically (with a recipe linking it to the thermal pressure), approximately accounting for the particles losses during the evolution.

Finally, 2D HD and MHD models have also been successfully used to investigate the formation of bow shock PWNe produced by fast moving pulsars escaped from their SNRs, and their variations depending on the properties of the ambient medium  \citep{Toropina:2001,Bucciantini_BowsII_2002, Bucciantini:2005,Olmi_Bucciantini_Morlino:2018,Toropina:2019}.
%
\begin{figure*}
\centering
	\includegraphics[width=.95\textwidth]{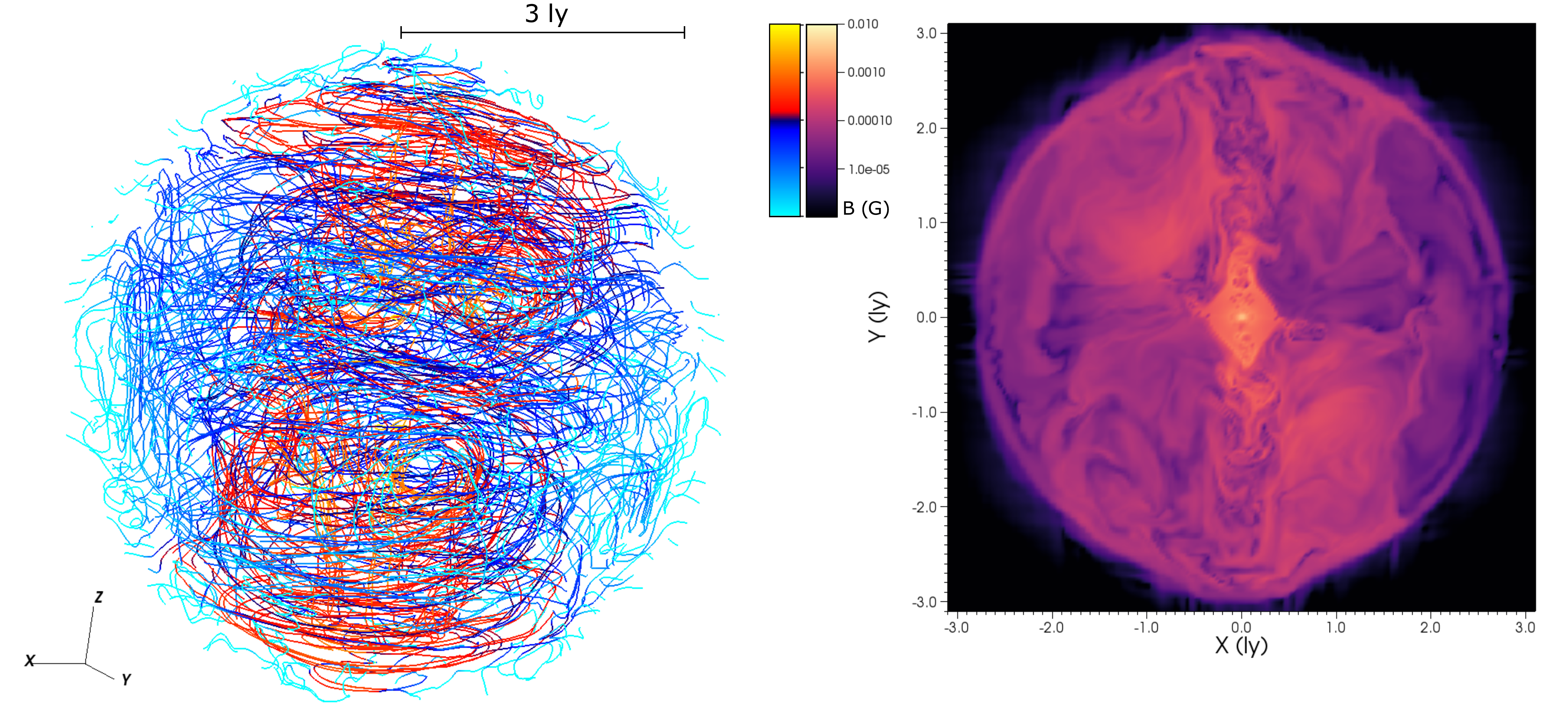}
    \caption{3D plot of the magnetic field lines (left side) and slice of the magnetic field intensity (right side) from the 3D simulation of the Crab nebula presented in \citet{Olmi:2016}, at a simulated age of $\sim 250$ yr. 
    Stream lines are colored with the intensity of the field as from the right panel, but using a different color scale to highlight the different structures. Both figures have been produced using VisIt \citep{childs2012visit}.}
    \label{fig:Bfield}
\end{figure*}
\subsection{3D models}
\label{sub:3Dmod}
With the first 3D MHD simulations of the Crab nebula \citep{Porth_sigma:2013} it became apparent that the solution to the sigma-paradox was finally in reach.
We have already seen how moving from 1D to 2D  permitted to gain more than one order of magnitude in wind magnetization.
Thanks to the development of efficient processes of magnetic dissipation in the nebula, a magnetization $\sigma\geq 1$ becomes finally accessible with 3D simulations, drastically reducing the impact of the sigma-paradox.
Configurations with a toroidal magnetic field are subject to current driven instabilities (kink-like, \citealt{Begelman:1998, Nalewajko:2012}), and signs of such a process have been detected in different PWNe \citep{Mori:2004,Pavlov:2003}.
Numerical simulations of the Crab nebula jet proved not only that the kink instability efficiently develops in the nebula, but that it is also responsible for the variations seen in the jets at different epochs \citep{Mignone_kink:2013}.
Mixing of the magnetic field induced by kink-like instabilities is so efficient that, even if the initial field configuration is fully toroidal, a poloidal component raises rapidly in 3D, becoming even dominant immediately outside the inner nebula (see e.g.  \citealt{Porth:2014,Olmi:2016}).
The mixing causes on one side the magnetic field geometry to become much more complex than in 2D, but also results in a more uniform distribution of the magnetic field in the nebular volume (see Fig.~\ref{fig:Bfield}).

Despite being the best way to account for the properties of a PWN, the use of 3D models is limited by the huge amount of resources (time/numerical) they need.
The spatial scales that must be reproduced are extremely different, from the injection region of the pulsar wind (smaller than the TS) to the PWN contact discontinuity, and the surrounding SNR bubble in case of young systems.
Moreover to correctly reproduce the pulsar wind, the injection site must be solved with a minimum number of grid cells ($\gtrsim 10-20$), and the injection region must always remain detached from the radius of the termination shock, otherwise the correct jump conditions at the shock cannot be ensured.
This translates in the necessity of a very high resolution at the center of the numerical domain, mapping a zone which represents only $\sim 1/100$ of the global size of the system.
The grid is then usually optimized with the use of an adaptive mesh technique, able to increase or decrease the resolution as needed, allowing to save a large amount of time, or with expanding grids.
Nevertheless the resources needed to run such models remain large and still prohibitive for long term evolution.
To date the longest 3D MHD simulation of the Crab nebula  reproduces $\sim1/4$ of the age of the source  \citep{Olmi:2016} and it required few millions of core/hours of computational time and several months to be run.
A longer evolution ($\sim7500$ yr) has been reached with HD simulations and expanding grids in the case of a high speed PWN ($v_{\mathrm{kick}}\simeq 300$ km s$^{-1}$) interacting with a composite SNR \citep{Kolb:2017}, producing a strongly  asymmetric system out of the reverberation phase.

A different approach has been applied to the modelling of bow shock pulsar wind nebulae.
For these systems, going 3D is particularly important to correctly capture the magnetic field topology and the development of turbulence in the tail, which are strongly affected by geometric constraints and relevant for the observational properties of the source.
A first attempt to 3D modelling of bow shocks, limited to the classical HD regime, was presented by \citet{Vigelius:2007}. 
Then, in recent years, the extension to the MHD relativistic regime was investigated by different groups \citep{Barkov:2019,Olmi_Bucciantini_2019_1}. In all these cases, the bow shocks have been modelled directly assuming the pulsar in interaction with the ISM, not considering then the transitional phase when the star is emerging from the SNR bubble. 
That transitional phase was indeed investigated in \citet{van-der-swaluw:2003}.

The PSR reference frame is generally used, with the star being positioned at a specific point of the computational domain. The star then sees the ambient medium as an incoming, cold flow, that might be magnetized or un-magnetized, depending in the specific case.
A large sample of different configurations has been investigated, varying the inclination of the magnetic field and pulsar-spin axis, the direction of the pulsar motion compared to the first two, the magnetization and  level of anisotropy of the pulsar wind.
The evolution is then followed for enough time to ensure the system dynamics has reached a relaxed quasi-steady regime.
Emitting properties can be computed using the same approach generally used for young systems. Particles responsible for the emission are injected at the wind termination shock with a broken power law distribution and their emissivity is computed as discussed in the following Sec.~\ref{sec:rad_and_acc}, with different possible assumptions on their density in the PWN (e.g. they can be considered as uniformly distributed or with a distribution shaped on the thermal pressure, see  \citealt{Olmi_Bucciantini_2019_2}). 

In a recent work \citep{Olmi_Bucciantini_2019_3} we have addressed the possibility for particles to escape the bow shock, modelling the evolution of the particle trajectories in the electric and magnetic fields of the MHD simulations.

%
\subsection{Highlights from 3D MHD models of young PWNe}
\label{sub:highl_young}
\begin{figure*}
\centering
	\includegraphics[width=.95\textwidth]{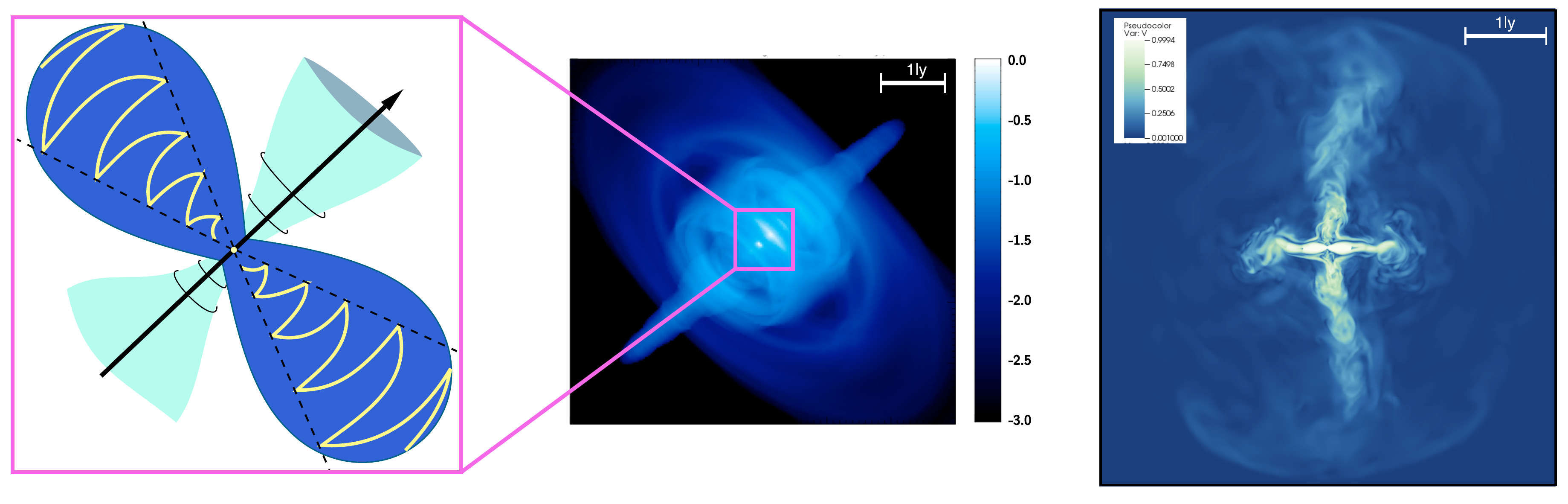}
    \caption{A selection of images for a PWN in the free-expansion phase, with maps from MHD simulations specific for the Crab nebula, elaborated from the simulations described in \citealt{Del-Zanna:2006} (middle panel --  2D) and \citealt{Olmi:2016} (panel on the right -- 3D). In particular, from left to right: toy model of the inner structure of the nebula, showing the oblate shape of the wind termination shock (the yellow spiral marks the striped wind) and the formation of jets due to hoop stresses at the polar front of the shock; 1 keV X-ray synthetic map of the Crab nebula, reproducing its inner morphology (map in logarithmic scale and intensity scaled to the maximum); intensity map of the 3D velocity field from a 3D simulation of the Crab nebula, with the evident formation of kinking jets.}
    \label{fig:HIGH3d_young}
\end{figure*}
%
Here we summarize the most important findings of 3D numerical simulations of young -- Crab like -- PWNe, according to \citet{Porth_sigma:2013, Porth:2014} and \citet{Olmi:2016}.
\begin{itemize}
    \item Solution of the sigma-paradox: values of the wind magnetization at injection larger than unity become possible thanks to the efficient magnetic dissipation produced in 3D. 
    Actually the average value of the magnetic field when the system has reached a self-similar evolution (at ages $\gtrsim 150$ yrs) is still a factor of $\sim1.5-2$ lower than the expected one (see Fig.~6 in \citealt{Olmi:2016}).
    \item Development of poloidal magnetic field: despite the magnetic field at injection is purely toroidal, turbulence and high-speed polar flows rapidly modify its topology. A polar component easily develops immediately beyond the inner nebula, and becomes even dominant in the outer nebula (see e.g. Fig.~11 in \citealt{Porth:2014} or Fig.~8 in \citealt{Olmi:2016}). 
    The complex structure of the magnetic field topology and its intensity can be seen in Fig.~\ref{fig:Bfield}.
    \item Despite the complex topology of the magnetic field, the magnetic pressure in the nebula is rather uniform on large scales (see Fig.~9 in \citealt{Olmi:2016}), as well as the total pressure (magnetic and thermal, see Fig.~3 in \citealt{Porth:2014}). This is a net difference with 2D MHD axisymmetric models, and explains why an HD approach seems more suitable to reproduce bulk and macroscopic properties of PWNe \citep{Olmi_Torres:2020}.
    \item The inner nebula is reasonably well described by 2D MHD axisymmetric models, including the Crab wisps and knot \citep{Komissarov:2004,Lyutikov:2016}. A direct comparison shows that, if one sticks to the inner -- toroidal -- nebula, the description obtained with 2D models does not differ much from the structure simulated in 3D (see Fig.~\ref{fig:HIGH3d_young}). This of course is a very important result, supporting the reliability of previous models in 2D and meaning that, if only interested in the properties of the inner nebula, one can safely use less demanding 2D simulations.
    \item Time variability of the inner nebula: the wisp-like variability close to the shock position is reproduced correctly and is comparable with what previously found with 2D models \citep{Camus:2009,Olmi:2015}. In 3D a number of variable non-axisymmetric structures is also found in the outer nebula, very similar to what observed at large scales in radio \citep{Bietenholz:2004}.
    \item Lacking of correct spectral description: despite the huge improvements in the description of the fluid structure, and in general in the global dynamics of the nebula, the emitting properties are not fully reproduced, both in terms of morphology and integrated spectrum. 
    The cause may be both in the -- still -- too low magnetic field on average, and/or in the approximated reconstruction -- a posteriori -- of the properties and history of the emitting particles.
\end{itemize}

\subsection{Highlights from 3D MHD models of bow shocks}
\label{sub:highl_bs}
\begin{figure}
\centering
	\includegraphics[width=.48\textwidth]{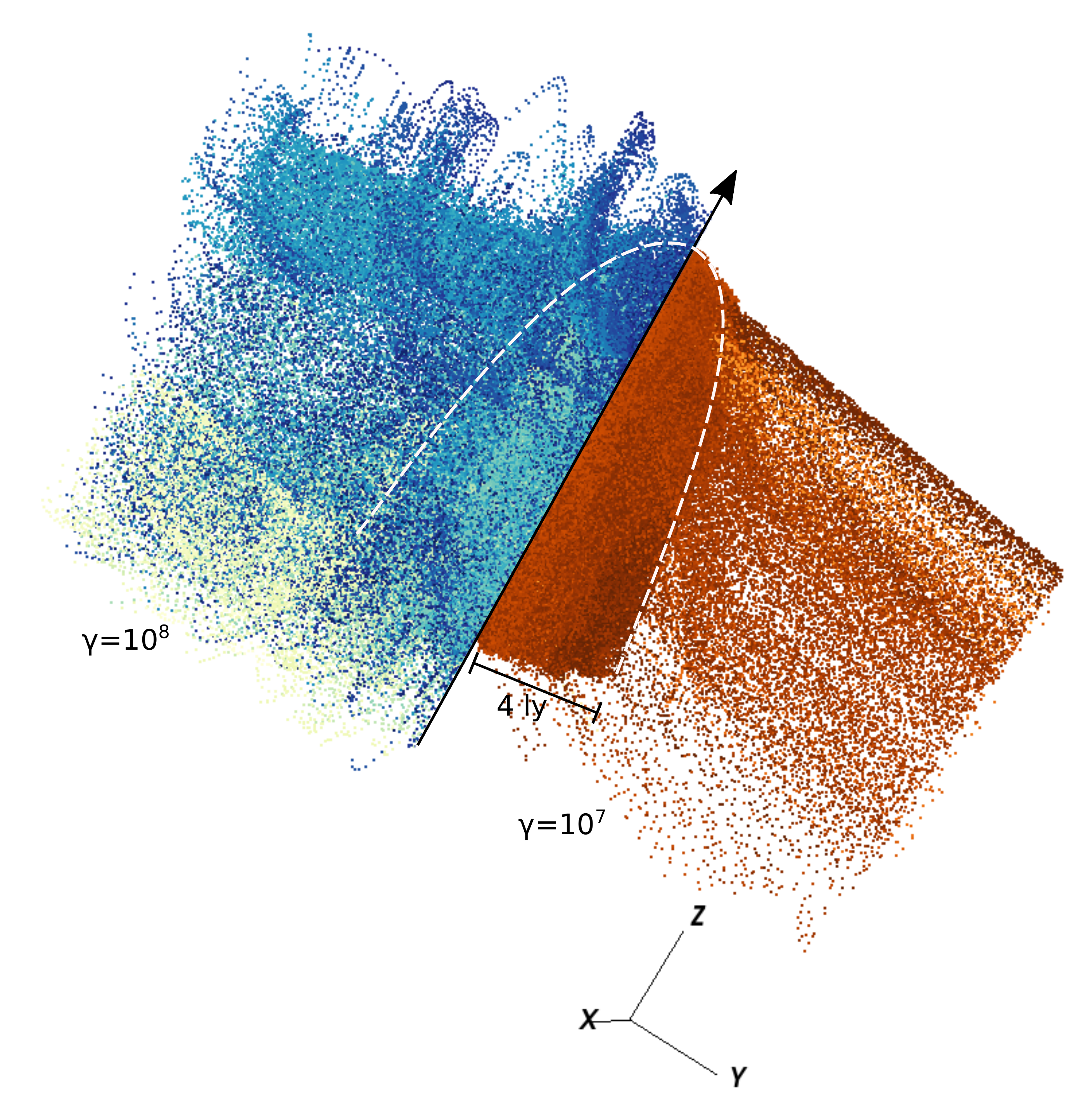}
    \caption{Comparison of two different regimes of particle escaping from a bow shock nebula from a 3D simulation: from massive and almost diffusive escape for high Lorentz factor particles ($\gamma=10^8$, left-blue colored side) to directional escape along external field lines of lower energy particles (with $\gamma=10^7$, right-orange colored side). Data come from the simulation presented in \citet{Olmi_Bucciantini_2019_3} and have been displayed using VisIt \citep{childs2012visit}. The inclination of the image is shown with the bottom triad, while the position of the bow shock is marked by the dashed white line.}
    \label{fig:bow+esc}
\end{figure}
%
In this subsection we summarize the results of 3D relativistic MHD simulations of bow shock nebulae. The discussion is mostly based on \citet{Barkov:2019} and  \citet{Olmi_Bucciantini_2019_1} for the dynamics, on \citet{Olmi_Bucciantini_2019_2} for the emission and polarization properties and on \citet{Olmi_Bucciantini_2019_3} for the properties of the particle escape.
\begin{itemize}
    \item The large scale structure of the bow shock is rather independent on the variation of the geometry and wind properties, with major deviations caused by different inclinations of the pulsar spin-axis and the direction of motion (maximum for $45^\circ$). The differences appear as small extrusions and blobs, in some cases resulting in periodic perturbation of the forward shock.
    \item The dynamics in case of an isotropic wind is very similar to the one found in 2D HD for intermediate values of the magnetization ($\sigma\simeq 0.1$).
    \item The dynamics in the tail is largely dominated by the level of magnetization and by the wind anisotropy: unlike anisotropy, the lower is the magnetization, the higher is the turbulence. Anisotropic models then are more turbulent than isotropic ones, showing strong mixing  also for large magnetization (see Fig.~3 of \citealt{Olmi_Bucciantini_2019_1}). In general, anisotropic and low magnetized systems are fully dominated by turbulence even close to the termination shock, resulting in the complete loss of information from the injection region along the tail. On the other hand, isotropic and highly magnetized systems show a coherent structure of the magnetic field, with the injection properties still affecting the dynamics far from the shock along the tail, in very good agreement with simplified semi-analytic models \citep{Bucciantini:2018}.
    \item No efficient magnetic amplification from turbulence. Even in presence of strong turbulence, the magnetic field tends to reach equipartition with the turbulent kinetic energy, in general smaller then the thermal energy in the tail. The only sign of field enhancement is found close to the contact discontinuity, possibly resulting from efficient shear instability amplification.
    \item Low turbulence cases -- emission and polarization properties:  a strong correlation between conditions at injection and surface brightness is found. The variety of observational morphologies is very wide: from bright heads to bright tails or, in some cases, bright wings. The polarization fraction is higher in the tail for higher magnetization.
    \item High turbulence cases -- emission and polarization properties: once magnetization drops (and anisotropy grows), turbulence starts to dominate also in the appearance of surface brightness, and distinguishing between the various cases becomes hard. When turbulence increases, the polarization fraction drops.
    \item The escape of particles from the bow shock is found to be an energy dependent process: the threshold for escape is set by the condition that the particle Larmor radius ($r_L$) in the equipartition magnetic field ($B\rs{eq}\sim \mathrm{few}\,\mu$G)  is equal to the typical size of the bow-shock in the head (the stand off distance $d_0$). 
    This translates in a Lorentz factor of the particles of: $\gamma\sim e B\rs{eq} \, d_0/(m_e c^2)$, that for typical systems\footnote{Namely:
    $d_0=10^{16}\, \mathrm{cm}\, \left[L_{36}/(\rho_1 v^2_{200})\right]^{1/2}$, with the luminosity expressed in units of $10^{36}$ erg s$^{-1}$, the ambient density in units of 1 proton per cm$^{3}$ and the PSR velocity in units of 200 km s$^{-1}$.} corresponds to $\gamma\gtrsim \mathrm{few}\times 10^7$, or in a particle energy $\gtrsim 10$ TeV.
    \item There is a transition in the escape process: particles manage to escape more easily if injected at the frontal polar region of the pulsar wind, while the others tend to remain confined in the tail. At lower energies, particles escape only in the presence of reconnection points at the magnetopause between the shocked pulsar wind and the ISM, and this might give rise to the appearance of one-sided jet-like features. Particles show an increasingly more diffusive escape with energy: the outflow becomes more uniform (see Fig.~\ref{fig:bow+esc}), but charge separation increases.

\end{itemize}

\section{RADIATION AND ACCELERATION}
\label{sec:rad_and_acc}
The pulsar is an excellent conductor. Charges inside it organize themselves in such a way that the internal electric field is screened.
But the electric field at the star surface is not, and it is strong enough to extract charged particles (leptons and possibly ions) from the star.
This generates a co-rotating magnetosphere that extends up to the star light cylinder $R_{LC}=cP$, with $P$ the pulsar period.
The magnetic field lines originating close to the pulsar magnetic axis (at the so called \textit{polar caps}) extend beyond the light cylinder and form  the open magnetosphere, through which the pulsar wind flows into the nebula. 
The rate at which particles are extracted at the PSR surface is given by \citet{Goldreich_Julian:1969}:
\begin{equation}\label{eq:GJrate}
    \dot{N}_{\mathrm{GJ}}=\frac{c\Phi}{e}\simeq 2.7 \times 10^{30} \left(\frac{B\rs{pc}}{10^{12}\,\mathrm{G}}\right)\left(\frac{P}{1 \, \mathrm{s}}\right)^{-2} \mathrm{s}^{-1},
\end{equation}
with $B\rs{pc}$ the magnetic field at the polar cap and $\Phi\simeq\sqrt{\dot{E}/c}$ the maximum potential drop between the pulsar and infinity.

Once extracted, particles are accelerated at  different locations along the open field lines, where they meet regions of un-screened electric potential.
As a consequence they emit high-energy photons that can be absorbed in the intense magnetic field surroundings the star, to generate electromagnetic cascades.
The number of pairs in the magnetosphere then increases by a large factor measured by the \textit{pair multiplicity} $k\sim 10^4-10^7$, namely the number of secondary leptons generated from the primary extracted from the star. 
The exact estimate of the multiplicity is very controversial \citep{2019ApJ...871...12T}, and there is a possibility to infer it from the modelling of the PWN properties.

Differently from leptons, ions cannot be generated in cascades, and so, if present, they must be a factor of $k$ less than pairs.
But given the difference in mass between ions and electrons ($m_p/m_e\sim 1800$), this does not necessary means that they are irrelevant in the PWN energetics (see e.g. \citealt{Amato_Olmi:2021} and the discussion therein).

PWNe reprocess a consistent part of the pulsar spin-down power into accelerated particles, with the Crab being the most efficient known at $\sim30\%$ efficiency. Only a very small fraction goes into pulsed radio to X-ray radiation ($\lesssim 1\%$), while a larger one might go into pulsed gamma-rays \citep{FermiLAT2:2013}. In general the study of the PWN emission is then relevant to obtain indirect information about pulsar physics.
PWNe shine at multi-wavelengths via non thermal emission. 
The primary emission mechanism, at least for a consistent period of their life (in free-expansion and possibly for large part of reverberation), is synchrotron radiation produced by the shocked wind particles interacting with the nebular, rather intense ($\sim 50-200\,\mu$G), magnetic field.
The synchrotron spectrum can be modelled as a set of broken power-laws. 
One (or two, for $t\gg\tau_0$, see \citealt{Pacini_Salvati:1973}) break is associated to synchrotron cooling ($E_c$). 
However, the others are typically thought to be associated with changes in the acceleration mechanism. The exact location of these breaks in the spectrum cannot be trivially inferred, and it is often based on more detailed modelling. In the Crab nebula, current models suggest that  the one between optical and X-rays is due to synchrotron cooling, while the one between radio and optical is attributed to a change in the particles acceleration mechanism. In other systems like MSH 15-5\textit{2} the two breaks are so close, and the spectral coverage so sparse, that it is hard to guess what is what \citep{2008ApJ...677..297N,Gaensler:1999,Gaensler:2002}. 

\subsection{Injection at the shock}\label{sub:injshock}
The particle injection spectrum is thus typically modeled as a broken power-law in the particle energy $E$:
\begin{equation}\label{eq:injf}
    Q(E,t)=Q_0(t)\left(\frac{E}{E_b}\right)^{-p_i} \,,   
\end{equation}
consisting of two distinct families, characterized by different injection indices $p_i$, at energies below or above the injection break ($E_b$).
The normalization function $Q_0(t)$ is determined by the requirement that the power injected in particles is a fixed fraction of the spin-down luminosity, namely: $(1-\eta) L(t) = \int_{E\rs{min}}^{E\rs{max}} Q(E,t) E \,dE$, where $\eta$ is the magnetic fraction (i.e. how much of the injection goes into magnetic energy), linked to the magnetization (Eq.~\ref{eq:sigma}) as: $\eta=\sigma/(\sigma+1)$.

A rather flat injection index ($p\rs{low}\sim 1-1.5$) is characteristic of the lower energy component, leading to a synchrotron spectrum with flux density $S_\nu(\nu)\propto \nu^{-\alpha\rs{low}}$ and spectral index  $\alpha\rs{low}\sim 0-0.3$.
The higher energy part indeed shows steeper spectra at injection ($p\rs{low}\sim 2-2.7$), leading to a synchrotron spectrum with $\alpha\rs{high}\sim 1-1.2$ or, as more commonly reported, a photon index $\Gamma=1+\alpha \sim 2-2.2$.
The recent PeV observations by LHAASO \citep{LHAASO_crab:2021,LHAASO_12s:2021} confirm, as originally suggested by \citet{Bucciantini:2011} and recently investigated by \citet{de-Ona-Wilhelmi:2022}, that high energy particles can be accelerated up to the pulsar voltage (see also \citealt{Khangulyan:2020}).

The possibility that particles responsible for the radio, optical and X-ray emissions belong to separate families, accelerated via different mechanisms, is supported also by the multi-wavelength variability observed in the Crab nebula.
The arc-like bright structures named \textit{wisps}, that appear very close to the TS location and move outwards with mild relativistic velocity, have been observed at multi-wavelengths  \citep{Hester:2002,Bietenholz:2001,Bietenholz:2004}, and shown to neither be spatially coincident nor characterized by the same velocity \citep{Schweizer:2013}.
In the MHD framework, which has  provided a very good description of the nebular dynamics, wisps trace the structure of the underlying plasma -- the magnetic field in particular -- and as such they can only be non-coincident if particles with different energies are produced at different locations of the shock.
This likely means that acceleration processes act differently in  different sectors of the shock \citep{Olmi:2015}.

The acceleration mechanisms proposed so far are mainly three: (i) diffusive shock acceleration, or Fermi-I like processes; (ii) diffuse acceleration due to stochastic magnetic reconnection, or Fermi-II, in MHD turbulence; (iii) acceleration conveyed by  driven magnetic reconnection.
Actually a fourth mechanism has been invoked: resonant absorption of ion cyclotron waves, which however requires the presence of ions in the wind, still not confirmed (nor excluded). 

Diffusive shock acceleration requires a very low magnetization to be effective ($\sigma \lesssim 10^{-3}$, \citealt{Sironi:2015}), a condition that can only be sustained at the equatorial sector of the shock, where the striped wind ensures a huge dissipation of the field, or very close to the polar axis, where the field naturally vanishes.
The power law index at injection of optical/X-ray emitting particles is compatible with what predicted by Fermi-I acceleration, and MHD models for the X-ray wisps are also in agreement with a scenario in which that particles are injected mainly at the equatorial front of the oblique termination shock.
On the other hand, driven magnetic reconnection requires a much higher magnetization ($\sigma \gtrsim 30$)  and a large pair multiplicity ($\kappa\gtrsim 10^8$ ), difficult to account within present models of pulsars magnetospheres \citep{Sironi:2011, 2019ApJ...871...12T}.
The power law index of the radio emitting particles at injection is on the other hand compatible with both reconnection and Fermi-II acceleration, while no particular information arises from wisps models in this case.
A possible conclusion is that radio particles are accelerated at higher latitudes along the shock, out of the equatorial sector, or via a more distributed acceleration in the nebula \citep{Olmi:2015,Lyutikov:2019}.

\subsection{Radiation mechanisms}\label{sub:radiation}
Once injected in the nebula, the particle energy $E$ evolves according to the following equation:
\begin{equation}\label{eq:part_evo}
    \frac{\partial E}{\partial t} = -  \frac{E}{R}\frac{\partial R}{\partial t} - c_2 E^2 \left( \frac{B^2}{8\pi}  + U\rs{rad}\right)\,,
\end{equation}
where $c_2=4/3 \,\sigma\rs{th}/(m_e^2 c^3)$, with $\sigma\rs{th}$ the Thomson cross section, and $U\rs{rad}$ is the energy density in radiation.
The first term on the right side of the equation represents the energy variation due to adiabatic expansion or contraction (depending on the evolutionary phase), while the second one describes radiation losses (synchrotron and IC).
From this equation one can easily find that the cooling energy $E_c$ is given in general by:
\begin{equation}\label{eq:Ecool}
    E_c = \frac{1}{R}\frac{\partial R}{\partial t}\frac{1}{c_2 \left( B^2/8\pi  + U\rs{rad}\right)}\,.
\end{equation}
Particles with $E<E_c$ are then dominated by adiabatic processes, losses or gains depending on the expansion or compression of the PWN. All particles with energy above $E_c$ are instead dominated by radiation losses.
In the free-expansion phase $E_c$ determines the cooling spectral break, that it is commonly found to move to higher energies with time \citep[e.g.][]{Pacini_Salvati:1973}.
During reverberation instead $E_c$ separates those particles that gain energy due to compression ($E<E_c$) from those that are still loosing energy due to radiation losses ($E>E_c$).

\begin{figure}
\centering
	\includegraphics[width=.48\textwidth]{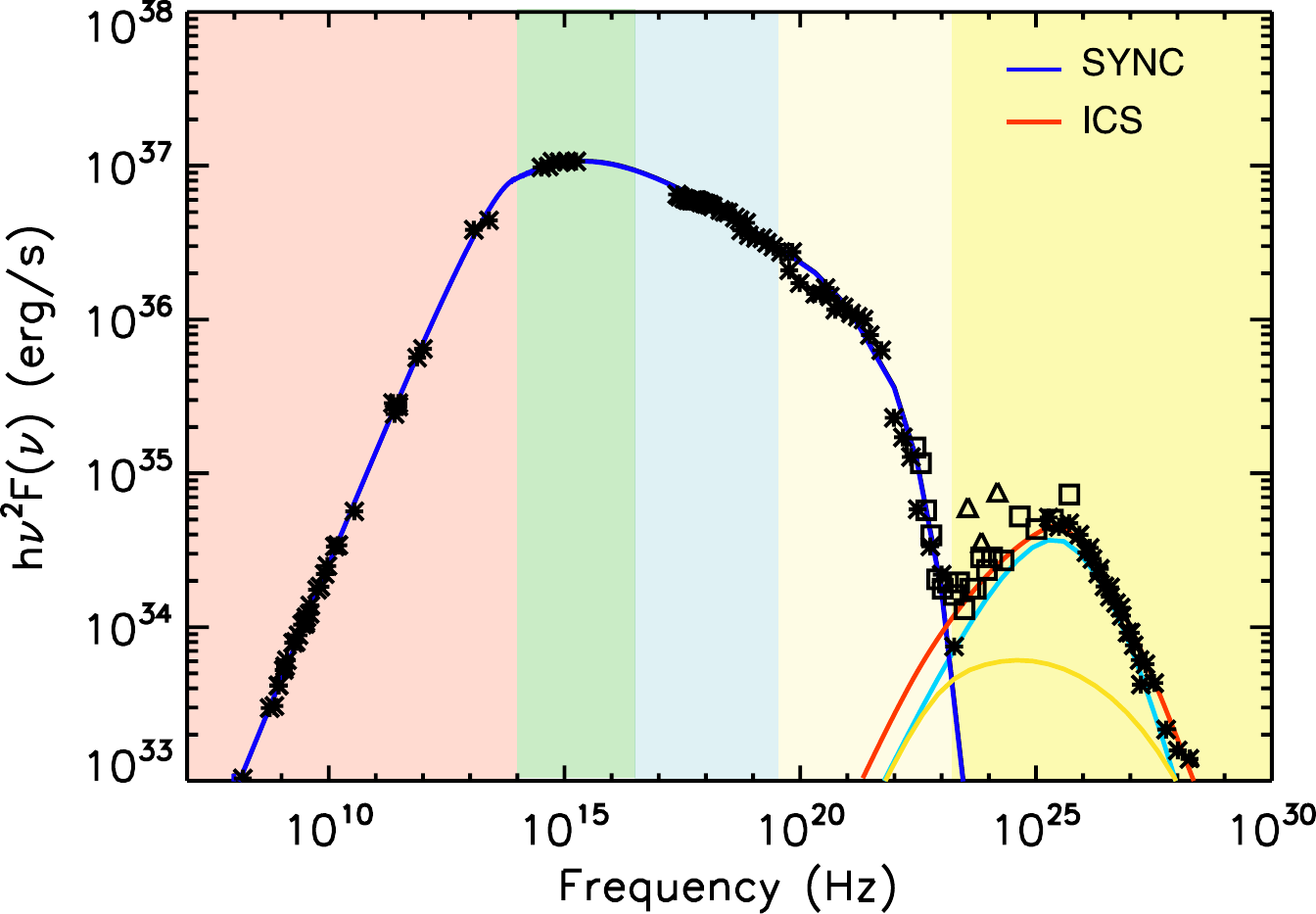}
    \caption{Non-thermal spectral energy distribution of the Crab nebula, elaborated from the original plot in \citet{Bucciantini:2011}, computed with a one-zone radiative model. Details of the assumed parameters and origin of data points can be found in the reference paper (see Fig.~1 and its caption).
    The different colored areas highlight the emission at various energy bands. From left to right: light-red for radio and IR; light green for optical and UV; light blue for X-rays (considering 0.1-100 keV); light yellow for the low energy gamma rays (up to the synchrotron limit $\sim 250$ MeV) and darker yellow for the high energy gamma-rays (fully due to IC). 
    The blue line is for the synchrotron component, the red one for the IC component, to which major contributions come from self-synchrotron Compton (in cyan) and scattering with CMB photons (in yellow).
    }
    \label{fig:SED_onezone}
\end{figure}

As already mentioned, the spectral energy distribution of a PWN is fully non-thermal, with the primary emitting mechanism, responsible for emission from radio up to few hundreds of MeV, being synchrotron radiation.
Higher energies are produced with the second emitting process characteristic of those systems: IC scattering between local photons and the same leptons responsible for the synchrotron emission.
An example of a full spectral energy distribution coming from a one-zone modelling can be found, in the specific case of the Crab nebula, in Fig.~\ref{fig:SED_onezone}.

The main contribution to IC in general comes from the interaction with the photons of the cosmic microwave background (CMB), with minor contributions from the interstellar radiation field and the synchrotron photons from the PWN itself.
An exception in this respect is the Crab nebula: due to its young age and intense magnetic field ($\sim 150-200\,\mu$G), its IC spectrum has in fact a consistent contribution from self-synchrotron radiation. %
The Crab nebula is moreover the only known case where the maximum energy of the accelerated particles is limited by radiation losses in the Galaxy, and is smaller than the maximum energy inferred by the available potential from the pulsar.

Once the injection spectrum has been defined (eq.~\ref{eq:injf}), the PWN luminosity can be computed as:
\begin{equation}\label{eq:Lpwn}
    L\rs{PWN}(\nu)=4\pi \int_{V\rs{PWN}} \left\{j_\nu^{\mathrm{SYNC}}(\nu) + j_\nu^{\mathrm{IC}}(\nu) \right\} dV\,,
\end{equation}
where $V\rs{PWN}$ is the PWN volume, while $j_\nu^{\,(i)}$ is the emissivity, obtained integrating over the particle distribution function either for  synchrotron or IC radiation as $j_\nu^{\,(i)}=\int_{E\rs{min}}^{E\rs{max}} \tilde{Q}(E,t) \, \mathcal{P}_\nu^{\,(i)}(E,\nu) \,dE$\,,
where $\tilde{Q}(E,t)$ is the evolved particle spectrum in the nebula, from the injection one defined in Eq.~\ref{eq:injf} (see e.g. \citealt{Bucciantini:2011}).
When computing the emissivity, in the 3D case, one should of course take also into account the spatial dependence due to orientation of the line of sight.
General expressions for the synchrotron and IC power $\mathcal{P}_\nu(E,\nu)$ can be found in many textbooks, e.g. \citet{Rybicki_Lightman:1979}.

\section{OBSERVING PWNe}
\label{sec:obs}
The firmly identified PWNe, with a detected associated pulsar, to date counts $\sim 60$ systems.
Most of them have been detected at X-rays thanks to Chandra \citep{Kargaltsev_Pavlov:2008}, while $\sim30$ are those detected also at gamma-rays with different instruments (see e.g. \citealt{Kargaltsev:2013} and the TeVCat catalog\footnote{The latest version of the TeVCat catalog con be found here: \url{tevcat2.uchicago.edu}} for an updated list). 
The number of PWNe detected up to now in the Galaxy  can be as high as $\sim90$  if
we also include those sources marked as putative PWNe from their spectral and morphological properties, but with no associated pulsar. 
Another factor of $\sim4$ is gained if we consider that a large part -- if not all -- of the present unidentified sources in gamma-ray surveys (e.g.  \citealt{HESScoll:2018-GPS}) are believed to be PWNe that have not been detected at lower energies, possibly due to their evolved stage and the consequent faint emission at lower energies.
If we consider a rate of birth of $10^{-2}$ pulsars yr$^{-1}$ in the Galaxy \citep{FGK:2006}, and an estimated lifetime of $10^5$ yr at gamma-rays, the expected number of detectable PWNe  at these energies in the Milky Way is in fact much larger than those really observed, with around $\sim 1000$ the expected systems.

\begin{figure}
\centering
	\includegraphics[width=.48\textwidth]{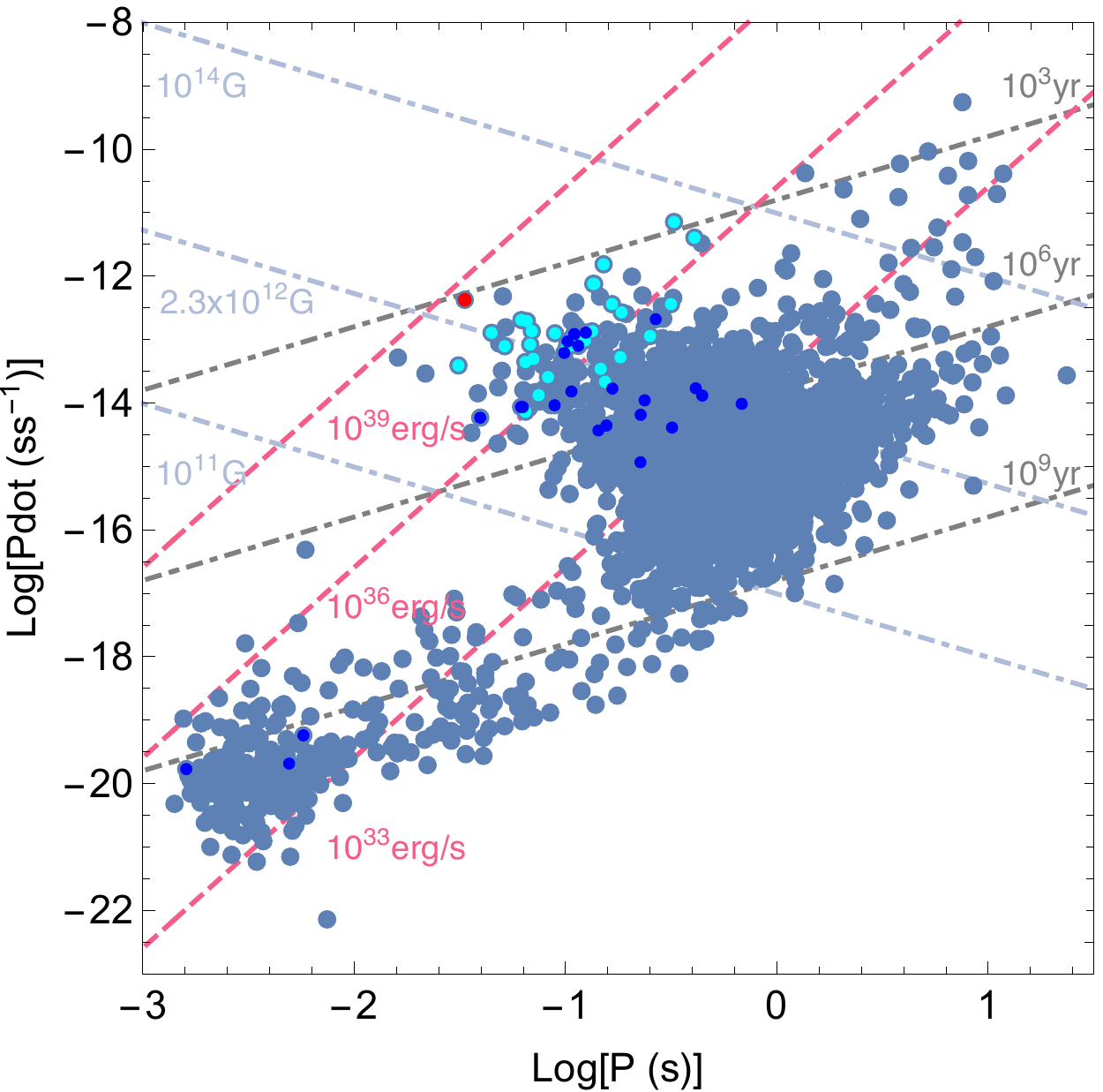}
    \caption{Distribution of the pulsars associated with identified PWNe on top of the complete pulsar population (in light blue) as taken from the ATNF catalog \citep{ATNFcat:2005}, version 1.67. 
    X-ray detected PWNe are shown as cyan circles or blue circles, in the last case those associated with fast moving pulsars.
    The Crab pulsar is shown as a red circle.
    All systems are given, respectively, in Table \ref{tab:cat_young} and \ref{tab:cat_old} of Appendix \ref{sec:cat}, with some useful parameters.
    For an easier interpretation of the plot we also give lines indicating the range of the surface magnetic field characteristic of pulsars associated with PWNe (in light blue, \citealt{Kargaltsev_Pavlov:2008}), in gray lines of characteristic age from $10^3$ yr to $10^9$ yr and in pink lines of fixed spin-down luminosity $10^{33}-10^{36}-10^{39}$ erg s$^{-1}$.}
    \label{fig:PPdotAll}
\end{figure}
%
%
In Fig.~\ref{fig:PPdotAll} we mark the position in the $P-\dot{P}$ diagram of the known PWNe (i.e. of their associated pulsar), to be compared with the total population of pulsars as derived from the ATNF catalogue \citep[][version 1.67, counting around 3300 stars]{ATNFcat:2005}.
It can be noticed that PWNe appear to be associated only with the youngest part of the pulsar population, with age ranging between a few hundreds of years to a million of years.
This can of course be partially due to a bias introduced by our inability to properly identify evolved systems, as in the case of the PWNe hidden in the population of unidentified gamma-ray sources, lacking of a multi-wavelength association.
The lifetime of a PWNe as an X-ray synchrotron nebula is in fact much smaller than its lifetime at very high energies. 
A 1 TeV photon can be produced via IC from the CMB -- or possibly from the IR background -- by an incoming electron of $\sim 10$ TeV energy, while the same photon requires a much more energetic particle to be produced via synchrotron radiation in the typical magnetic field of a PWN.
In fact, even considering a very low magnetic field of $\sim 10\mu$G, characteristic of evolved systems, a 50 TeV electron is necessary to produce a 1 keV photon.

The lifetime of a lepton of energy $E_{\mathrm{e, TeV}}$, expressed in units of TeV, against synchrotron losses in a magnetic field $B_{\mu\mathrm{G}}$, in units of $\mu$G, is in fact given by:
\begin{equation}\label{eq:tau_synchE}
    \tau\rs{synch}\simeq 25 \, \left( \frac{B_{\mu\mathrm{G}}}{100} \right)^{-2} \left( \frac{E_{\mathrm{e, TeV}}}{50}\right)^{-1}\mathrm{yr}\,.
\end{equation}
It is then clear that the more energetic the lepton is, the quicker it radiates its energy away in the form of synchrotron emission, and the less long it survives. 
The same can be seen if looking instead at the energy of the synchrotron emitted photons (in keV units):
\begin{equation}\label{eq:tau_synchPH}
    \tau\rs{synch}\simeq 55.2 \, \left( \frac{B_{\mu\mathrm{G}}}{100} \right)^{-3/2} \left( \frac{E\rs{ph, keV}}{1}\right)^{-1/2}\mathrm{yr}\,.
\end{equation}
This makes a PWN detectable at X-rays only for a limited fraction of its life, when the pulsar is still powerful enough. 
On the contrary a PWN shines at radio energies for longer time, being the lifetime of radio emitting electrons much longer. These are also the same particles responsible for the long-living IC gamma-ray emission.
As discussed previously, the detection at radio frequencies, especially for evolved, extended or diffused systems, might be difficult for multiple reasons, first of all instrumental limitations.
Old nebulae are then likely to be detected mainly at gamma-rays, where we still lack in resolution, and their morphology is then difficult to be determined.
A high level of linear polarization is one of the key properties of synchrotron emission \citep{Westfold:1959,Legg_Westfold:1968}. For the typical particles distribution functions that are observed in PWNe, the polarized fraction theoretically can be as high as 70\%. It was indeed thanks to its high optical polarization, that synchrotron emission was recognized for the first time as the main emission mechanism in an astrophysical source, the Crab nebula \citep{Baade:1956,Oort_Walraven:1956, Woltjer:1958,Velusamy:1985}.

Polarization is customarily observed in radio, and  maps are available for many PWNe. The naive expectation is that the polarized structure in PWNe should correspond to a mostly toroidal magnetic field, as the one generated by a fast spinning rotator. There are indeed a few systems like Vela \citep{2003MNRAS.343..116D} and G106.6+29 \citep{2006ApJ...638..225K} where a well defined large scale toroidal pattern is observed with polarized fraction as high as 30\%-40\%. 
However, there is a wide variety in the radio polarization structures: some systems show a large scale radial/dipolar pattern \citep{2008ApJ...687..516K,2022ApJ...930....1L}; while others have more random one, like the Crab \citep{1990ApJ...357L..13B,Aumont:2010}, with little to no correlation with respect to bright emission features. Polarization is also available for old systems \citep{2016ApJ...820..100M} and for a handful of bow-shocks \citep{Ng:2010,2016ApJ...820..100M}.
Being radio emission in young or middle aged PWNe dominated by the outer regions (since radio emitting particles are older and  then fill the entire nebula), more subject to the interaction with the environment, radio polarized measures provide at best a good estimate of the degree of ordered versus disordered magnetic field for the overall nebula, but cannot be used to investigate the conditions in the inner regions, where particle acceleration takes place.

Optical and near IR polarization is only available for three systems: the Crab \citep{Hester:2008,Moran:2013} where, due to the presence of a large foreground, only the bright knot and wisps have been studied and show a high level of polarization of 40\%-50\%, compatible with a toroidal magnetic field; G21.5-0.9 \citep{2012A&A...542A..12Z}, where a small internal torus is observed with polarized fraction as high as 50\%; SNR 0540-69 \citep{Lundqvist:2011}.

Until the launch of the Imaging X-ray Polarimetry Explorer in December 2021 \citep[IXPE, ][]{IXPE:2022}, the Crab nebula was the only PWNe (in-fact the only astrophysical object) to have a measured X-ray polarization \citep{Weisskopf:1978}. 
The polarized fraction was found to be 19\%, with a polarized angle marginally compatible with the symmetry axis inferred from fitting the X-ray torus \citep{Ng_Romani:2004}. 
With IXPE an X-ray spatially resolved polarized measure finally became available for Crab \citep{IXPE_Crab:2022}, Vela, and MSH 15-5\textit{2}, while integrated polarimetry will also be measured in a handful of other PWNe.

In Appendix \ref{sec:cat} an updated ``catalog'' of all the Galactic PWNe with an associated PSR known at present, divided in low and high speed systems, is reported in the two tables, with some ancillary information.

\subsection{Young systems}
\label{subsec:young_systems}
To date we have identified less than 20 sources as PWNe in their free-expansion phase. 
They constitute a large part of the catalog of the PWNe detected mainly at X-rays and not associated with a fast moving pulsar ($\sim 40$ sources), reported in Table~\ref{tab:cat_young}.
%
\begin{figure*}
\centering
	\includegraphics[width=.95\textwidth]{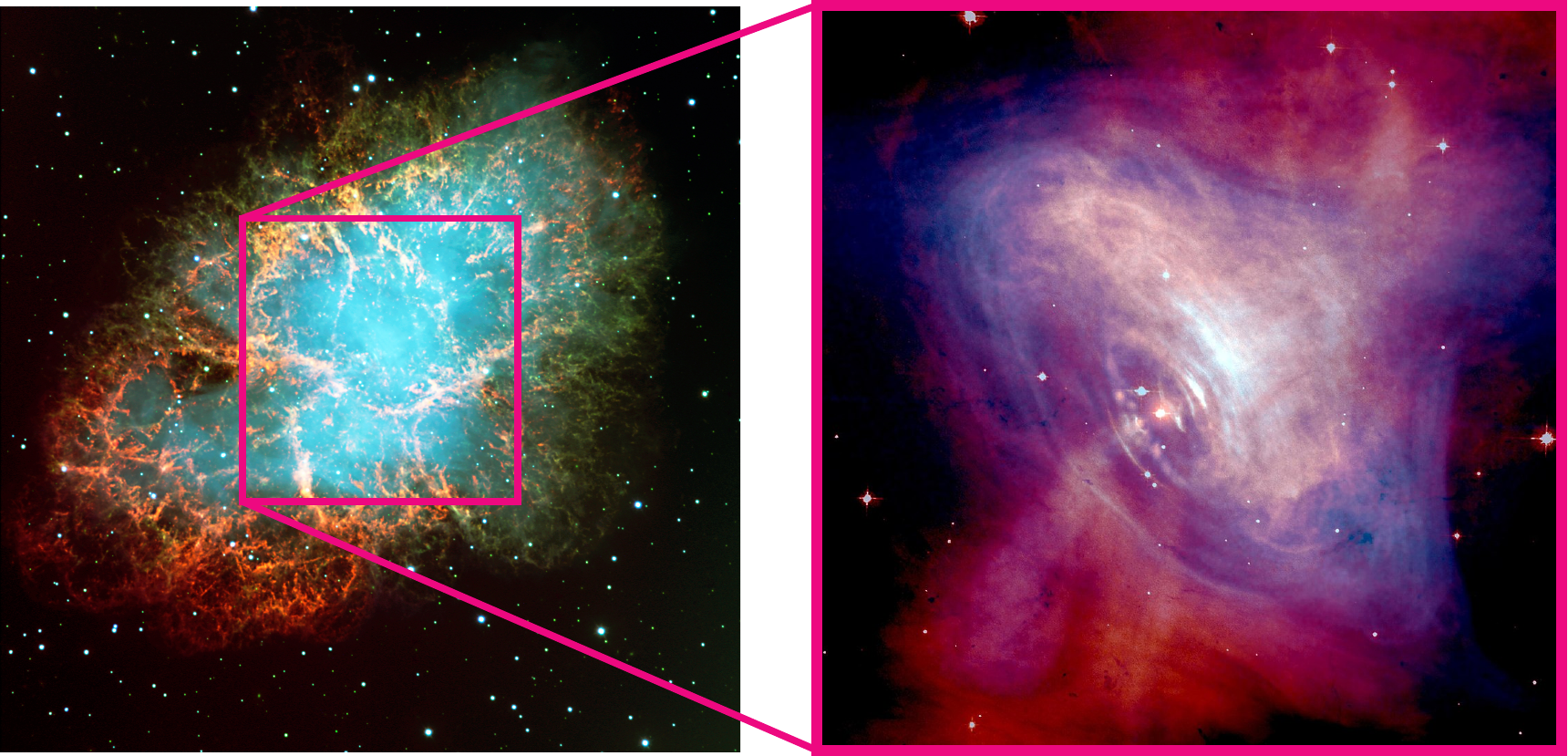}
    \caption{Left panel: Composite optical image of the Crab nebula, \textit{Credits: ESO}. Right panel: Combined optical (in red -- from Hubble) and X-ray (in blue --  from Chandra) images of the Crab nebula, \textit{Credits: Optical -- NASA/HST/ASU/J. Hester et al. ; X-Ray -- NASA/CXC/ASU/J. Hester et al.}}
    \label{fig:Crab_cfr}
\end{figure*}
%
%
From the morphological point of view, synchrotron dominated systems are characterized by a larger extension at lower energies than at higher ones (see e.g. Fig.~\ref{fig:Crab_cfr}), with sort of a spherical/elliptical shape. 
X-rays highlight the inner nebula, revealing the presence of a jet-torus structure (see the zoom in of Fig.~\ref{fig:Crab_cfr}), believed to be a rather common feature in young PWNe, first discovered in the Crab and successively identified in another bunch of systems.

Another common feature of the observed torii is the appearance of enhanced brightness at one side (effect of the relativistic Doppler boosting of the emitting particles moving towards the observer) and the presence of variable, both in brightness and position, arc-like (the wisps, \citealt{Scargle:1969,Bietenholz:1991,Bietenholz:2001,Helfand:2001,Pavlov:2001,Bietenholz:2004}) or point like (\textit{knots}, \citealt{Lou:1998}) structures marking the high variability of the inner nebula, where (most of?) the particles are accelerated.
As mentioned in section \ref{sub:2Dmod}, the discovery of the complex inner structure of the Crab nebula was what prompted the move from 1D models to 2D MHD simulations.

As discussed previously (Sec.~\ref{sec:rad_and_acc}), young PWNe are characterized by extremely broad band spectra, extending from radio to gamma-rays; with the advent of LHAASO, now the Crab spectrum has been further extended above PeV energies \citep{LHAASO_crab:2021}.

\subsection{Middle aged -- reverberating systems}
\label{subsec:mid_systems}
\begin{figure}
\centering
	\includegraphics[width=.45\textwidth]{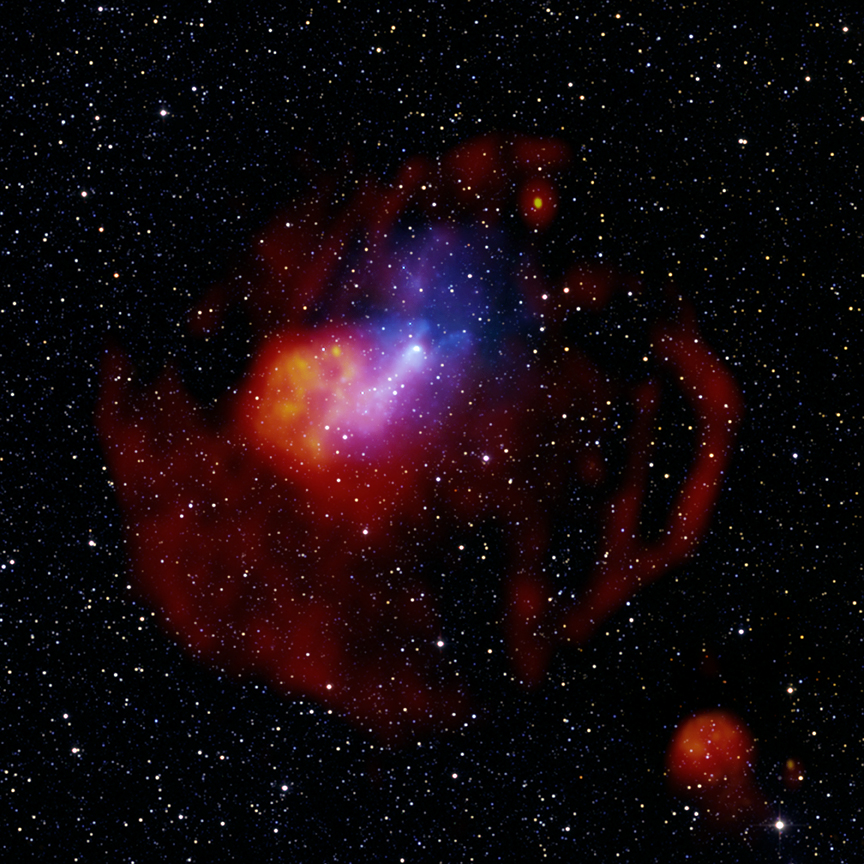}
    \caption{Composite IR (for the stellar field), radio (in red color) and X-ray (in blue) image of the PWN G327.1-1.1, one of the very few systems in clear interaction with the SNR reverse shock. 
    \textit{Credits:  X-Ray -- NASA/CXC/SAO/T. Temim et al. and ESA/XMM-Newton; Radio: SIFA/MOST and CSIRO/ATNF/ATCA; IR: UMass/IPAC-Caltech/NASA/NSF/2MASS.}}
    \label{fig:G327}
\end{figure}

The reverberation phase, characterized by the interaction with the SNR reverse shock, is hardly identifiable. At present we only know a handful of systems showing clear evidence of being in that stage, among which Vela X \citep{Blondin:2001}, the Boomerang nebula \citep{Kothes_boomerang:2006} and the Snail in G327.1-1.1 \citep[][shown in Fig.~\ref{fig:G327}]{Temim:2009, Temim:2015}. 

Independently of the pulsar speed, middle aged systems are expected to show large asymmetries.
The interaction with the SNR reverse shock is not expected to happen spherically, as simplified one-zone models are forced to assume (see Sec.~\ref{sub:0Dmod}). The compactness of the PWN contact discontinuity itself is partially destroyed by R-T like instabilities (Sec.~\ref{sub:postrev}) well before the onset of reverberation. 
Of course, in case of a high proper motion of the star, the asymmetry is even larger.
The asymmetry introduced in the PWN morphology in this stage is then expected to be a common feature of almost all middle aged systems, as well as of old ones, if they have not become bow shock nebulae. The original PWN bubble might be fragmented, with radio (and gamma-ray) separated bubbles surrounding the remaining nebula and expanding under adiabatic forces. The background of the PWN can be then very noisy, making it difficult to be identified at radio or gamma-rays. On the other hand, X-ray emission might simply be too faint.

Estimating the level of fragmentation and mixing of the PWN after reverberation can be quite complex: HD simulations typically do not converge, because in the HD regime the fastest growing scale of R-T like instabilities is set purely by numerical viscosity at the grid scale  \citep{Blondin:2001,Bucciantini:2005}, while MHD simulations have not been performed for old systems; observationally the mixing has only been estimated for the Snail \citep{2016ApJ...820..100M}, suggesting a pulsar wind filling factor of order of 50\%. For a discussion of the possible impact of mixing on thin-shell modelling see \citet{Bucciantini:2011}.

\subsection{Bow shock nebulae}
\label{subsec:mid_systems}
Through the combination of radio, H$_\alpha$ (in case of a partially ionized ambient medium) and, especially, X-ray observations, nowadays we have identified 25 fast moving pulsars with an associated bow shock nebula.
They are listed in Table~\ref{tab:cat_old} and are marked with blue colored circles in Fig.~\ref{fig:PPdotAll}.

Few bow shock nebulae show an elongated X-ray tail, in some cases associated with an even more extended radio tail. 
Only a part of them shows a spectral variation along the X-ray tail, in particular a softening indicating synchrotron cooling (e.g. the Mouse and the Lighthouse, \citealt{Kargaltsev:2017}).
To date no TeV emission has been detected from bow shocks directly, while UV has only been detected in correspondence with H$_\alpha$ emission, probably coming from the heated shocked ISM.
The bow shock head appears not to have a standard morphology, with even drastic variations from one object to another. 
As  shown by 3D MHD models \citep{Barkov:2019,Olmi_Bucciantini_2019_1,Olmi_Bucciantini_2019_2}, these variations can be ascribed to intrinsic differences in the geometry of the pulsar magnetosphere, in the orientation of the pulsar spin-axis with respect to the pulsar direction of motion, and the orientation with respect to the observer's line of sight.

\begin{figure}
\centering
	\includegraphics[width=.45\textwidth]{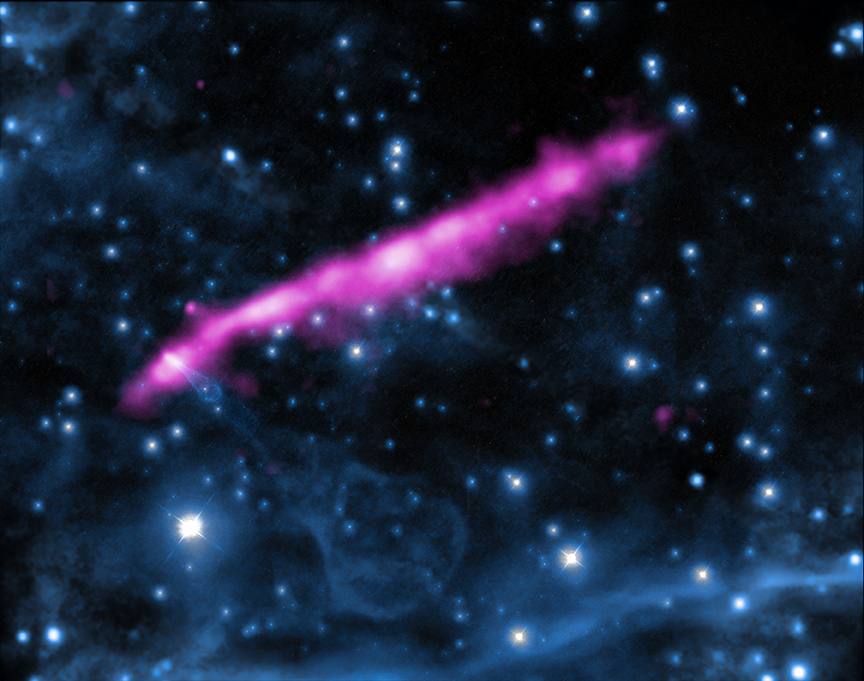}
    \caption{Composite optical (in blue color) and X-ray (in magenta) image of the Guitar BSPWN, generated by the fast moving pulsar J2225+6535, and its extended X-ray misaligned tail. 
    The Guitar nebula is also one of the systems characterized by the modification of the tail structure in the so called \textit{head-and-shoulder} shape, possibly indicating mass loading from the ambient medium into the tail.
    \textit{Credits:  X-Ray -- NASA/CXC/UMass/S.Johnson et al; Optical: NASA/STScI \& Palomar Observatory 5-m Hale Telescope.}}
    \label{fig:Guitar}
\end{figure}

In some cases the structure of the bow shock appears to be modified in the so called  \textit{head-and-shoulder} shape: the bow shock shows an  evident widening with distance from the pulsar, with possibly a periodic structure, as the famous example of the Guitar nebula \citep[][see e.g. Fig.~\ref{fig:Guitar}]{Chatterjee:2004,van_Kerkwijk:2008}.
This is believed to be the sign of the mass loading of ambient neutral atoms into the bow shock through the shocked ISM \citep{Morlino:2015}; those atoms then interact with the pulsar wind and modify its dynamics \citep{Bucciantini_bowsI:2001, Bucciantini_BowsII_2002}. This effect has been proved through numerical simulations by \citet{Olmi_Bucciantini_Morlino:2018}, showing that the lateral expansion of the bow shock tail is a function of the pulsar Mach number only, namely it increases with the Mach number as the effect of the augmented ram pressure exerted by the ISM on the bow shock nebula contact discontinuity.

In recent years bow shock nebulae have gained renewed interest thanks to the detection of collimated, extended and generally highly misaligned (with respect to the direction of motion),  jet-like features, only visible at X-rays (observed in the Chandra band: 0.5-8 keV), usually referred to as \textit{misaligned tails} \citep{deLuca:2011,Pavan:2014,Klingler:2016,deVries:2020,WangD:2021,deVries:2022,deVries_Romani:2022}.
These structures were already observed few years ago surrounding a couple of systems (e.g. the Lighthouse nebula), but with the recently increased number of detection  they now seem a rather common feature of these evolved nebulae.
At present a generally accepted interpretation \citep{Bandiera:2008} is that they are produced by high energy particles (close to the maximum limit of the potential drop) leaking from the bow shock nebula, and then producing emission via synchrotron radiation in the local magnetic field.
The observed asymmetry appears to be related to the mutual inclination of the spin-axis of the pulsar, its magnetic field, the pulsar speed and the direction of the magnetic field lines in the ambient medium \citep{Olmi_Bucciantini_2019_3}.
Indeed an open question is how to amplify the magnetic field from the ISM value to the order of few tens of $\mu$G required to produce the observed emission. 
%
\begin{figure}
\centering
	\includegraphics[width=.45\textwidth]{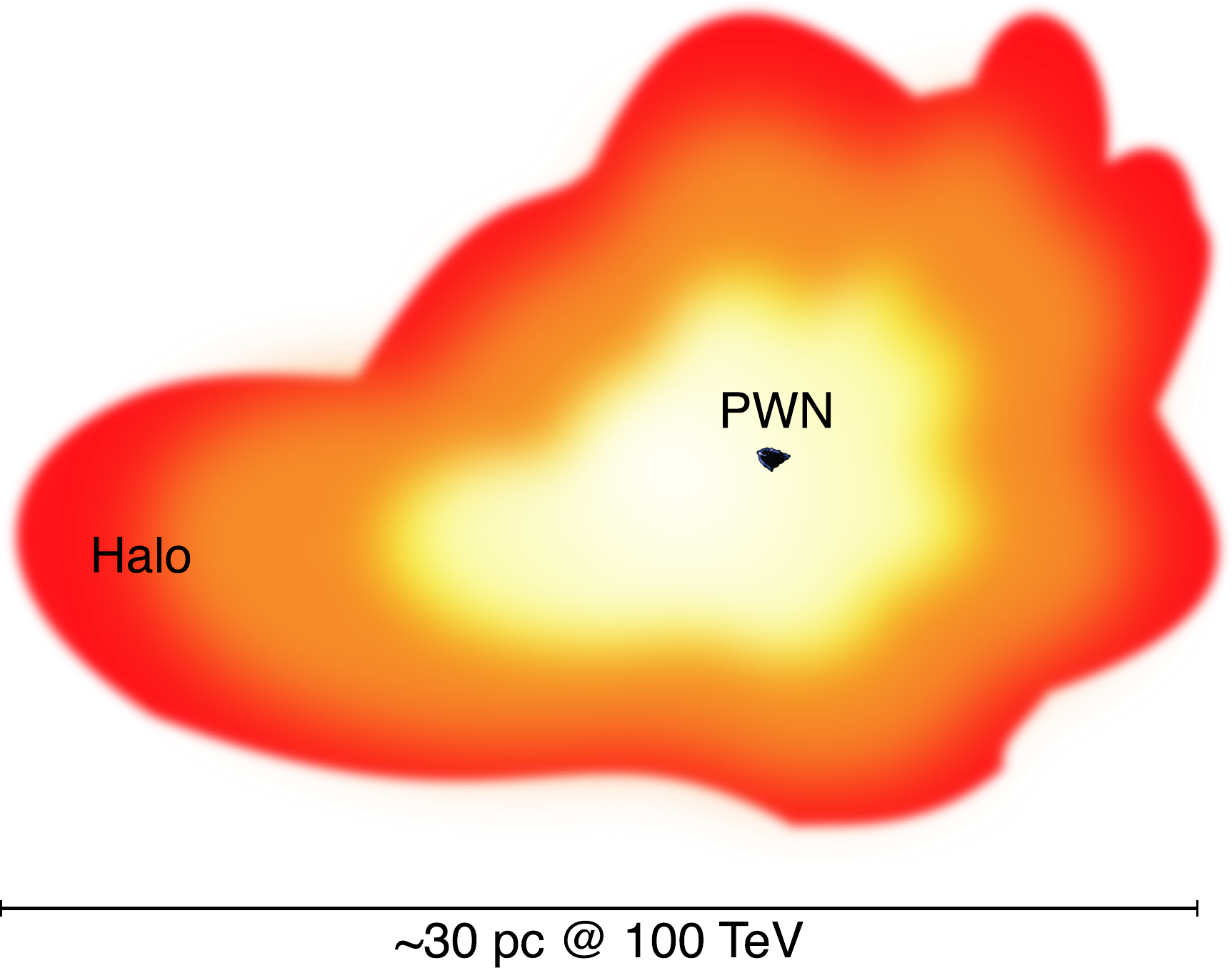}
    \caption{Sketch roughly comparing the size of the TeV halo around the Geminga pulsar wind nebula (from HAWC measures at 100 TeV) and that of the X-ray pulsar wind nebula (image adapted from the original composite picture at X-rays -- Chandra -- and IR -- Spitzer).
    \textit{Credits for the PWN map:  X-Ray -- NASA/CXC/PSU/B. Posselt et al; IR: NASA/JPL-Caltech}.}
    \label{fig:Geminga}
\end{figure}

Evolved pulsars are also associated with the formation of TeV halos \citep{Abeysekara:2017}, that have been again interpreted as high-energy particles escaping from the bow shock nebula, and then diffusing (with some suppression) in the ambient medium to form bright gamma-ray bubbles, much more extended than the original bow shock (see e.g. a sketch for the case of Geminga in Fig.~\ref{fig:Geminga}).

\section{Where we are and WHERE WE GO}
\label{sec:open_and_where}
\subsection{Observational prospects}
\label{sub:obs_future}
The actual population of PWNe is expected to increase largely in the next future, especially thanks to very high energy ($>100$ GeV) observations.
The actual Galactic plane survey from the  H.E.S.S. telescope \citep{HESScoll:2018-GPS} found that more than half of the 24 extended sources detected are identifiable with PWNe (14, mainly thanks to their multi-wavelength counterpart). 
In the Fermi-LAT 3FGL catalog \citep{3FGL:2013} the unidentified sources are $\sim 20\%$ of the total and it is plausible that most (all?) of them are actually PWNe with no direct association with a known pulsar.
As we already said, thanks to their longer lifetime as gamma-ray emitters, middle aged PWNe will likely dominate the very high energy sky, possibly representing up to $\sim60\%$ of the Galactic sources, and a huge number of new detection ($\sim 200$) may be expected with the upcoming Cherenkov Telescope Array, thanks to its unprecedented angular and energy resolution \citep{Remy:2022, Fiori:2022}. 
A very challenging problem will be that of the source confusion in the crowded Galactic plane, that will reduce the number of identified sources with respect to theoretical estimate \citep{Mestre:2022}.
Moreover an additional source of confusion might be that of TeV halos associated with evolved pulsars. 
Given their extended and weak emission, they are difficult to identify, and likely constitute a background noise for other sources. 
However the number of expected halos in the Galaxy is still matter of debate, ranging from few hundreds to a few, depending on their physical interpretation \citep{Sudoh:2019,Giacinti:2020,Martin:2022}.

A very exciting recent result is the detection of PeV photons coming from 12 sources in the Galaxy, plus the Crab nebula, by \citet{LHAASO_crab:2021,LHAASO_12s:2021}.
Unfortunately, the limited angular resolution of LHAASO does not permit to identify the exact location of the source (and origin of these energetic photons), except for the case of the Crab.
One or more pulsars can be found in the same region covered by the PSF of the instrument for all the 12 sources. Out of these, 11 result to be theoretically compatible with being powered by a pulsar \citep{de-Ona-Wilhelmi:2022}, meaning that pulsars and their nebulae might also be the most numerous class of extremely high energy emitters in the Galaxy.

PeV data are also fundamental to either confirm or exclude the presence of an hadronic component in the pulsar wind  \citep{Atoyan:1996,Bednarek:1997,Bednarek:2003,Amato:2003}.
If present, hadrons might show up only at the very high energies, where the leptonic emission from IC drastically falls due to Klein-Nishina suppression \citep{Amato_Olmi:2021}, thus the next generation of imaging atmospheric Cherenkov telescopes will have a crucial role in answering this question.

The recently launched IXPE satellite \citep{IXPE:2022}, operating in the range 2-8 keV, would enable us to sample and image, for the first time, the magnetic field structure (not just its strength) and the level of turbulence in a handful of PWNe, resolving the magnetic field pattern in the central region of the torus-arcs characteristic of young objects, and possible in the jets. The three main targets of the mission for space resolved polarimetry are: the Crab nebula, Vela and MSH 15-5\textit{2}. Other PWNe will be observed in the second and third year of operations. 
Prior to IXPE  the Crab nebula was the only object to have been observed with a detected average X-ray polarization of 19\%. IXPE will open a new observational window into the way we understand and characterize these objects. Preliminary results are coming in these same days \citep[e.g][]{IXPE_Crab:2022}. By the end of the mission we expect to have a much better understanding of the dynamical conditions in these relativistic accelerators.

With the approaching end of operations of the Chandra telescope, a very important instrument for future observations of PWNe will be the LYNX X-ray  observatory\footnote{\url{www.lynxobservatory.com}}, that thanks to its improved sensitivity and field of view promises to open new windows of opportunity to investigate the structure and details of X-ray sources (an example is the possibility to finally detect the compact object in SNR 1987A, see e.g. \citealt{Greco:2021}).

\subsection{Modelling prospects}\label{sub:mod_future}
Despite the astonishing progresses made in the last two decades, there are still many important open questions in our understanding of the physical processes operating in PWNe.
If a part of the historical open problems seem nowadays to be solved, as the case of the long standing sigma-paradox, some are still not properly answered, among which:
\begin{itemize}
    \item What are the physical mechanisms responsible for particle acceleration at the different energies? All the proposed mechanisms have strengths and weaknesses and require precise -- and very diverse -- physical properties of the nebular plasma to be viable.
    \item What is the origin of the gamma-ray flares in the Crab nebula? And is the Crab the unique source producing flares? Many possibilities had been proposed and investigated, most of them requiring a mG magnetic field at the emitting region (this is the case for powerful reconnection events in $\sigma\gg 1$ regions), much larger than that estimated from PeV emission. A very detailed discussion of this point, and of the proposed models, has been recently reported in \citet{Amato_Olmi:2021}.
    \item Are hadrons present in the pulsar wind? Despite being certainly a minority by number, hadrons could even be -- if present -- energetically dominant in the wind, changing completely our understanding of its properties.
\end{itemize}
In the latest years new questions have been added, especially thanks to observations of evolved systems that have revealed a number of unexpected features:
\begin{itemize}
    \item How and with what efficiency particles can escape from evolved pulsar wind nebulae? 
    \item How escaped particles produce misaligned tails or extended halos?
    \item Do pulsars accelerate particles at the theoretical limit of the maximum potential drop?
    \item How does the interaction with the reverse shock of the SNR modify the emitting properties and morphology of evolved PWNe? 
    \item Which ingredients of current modelling need to be modified/improved to interpret the upcoming large amount of gamma-ray observations and to manage source confusion? Is there a way to theoretically model the possible different morphologies and spectra of evolved PWNe that can help in their identification, especially in the lack of a multi-wavelength counterpart?
\end{itemize}
Recent modelling efforts go in the direction of trying to answer these questions. 
How particles escape from evolved systems seems now to be clarified, as discussed in \citealt{Olmi_Bucciantini_2019_3}. The efficiency of the escape process is strictly linked to the particle energy, with only the most energetic particles, close to the maximum energy achievable, able to escape in large fractions.
The possibility for them to be revealed as extended and diffuse, or asymmetric and thin, structures is also partially accounted for with numerical models, that predict a variety of escaping processes depending on the properties of the system and those of the surrounding medium. 

We still lack an understanding of is what happens once particles have been injected in the ambient medium. What process causes the reduction of the Galactic diffusion length  in the vicinity of pulsars (if this is the case), to produce TeV halos with the observed size through diffusion? 
And also, what amplifies the magnetic field to the value 
needed for the observed synchrotron emission from misaligned X-ray tails, a factor of 2 to 10 larger than that expected in the ISM?
To answer these questions new dedicated modelling of particle propagation must be investigated, considering diffusion properties, development of self-turbulence and the onset of instabilities able to modify the properties of the ISM, possibly using a hybrid approach between pure MHD and  Particle In Cell (PIC) techniques, that correctly accounts for the evolution of particles in the plasma.

A better understanding of the composition, and of the physical properties, of the pulsar wind passes through a correct interpretation of high energy gamma-ray data. In particular, the firm identification of the observed PeVatrons will shed light on the possibility for pulsars to efficiently accelerate particles very close to the theoretical limit. A refined modelling of the spectral properties through evolutionary phases will also be extremely important.

A very challenging point will be to develop models that help in disentangling the PWNe contribution to the gamma-ray Galactic emission from that of other sources. Part of this requires to understand how to model evolved systems, both in terms of emission and morphology, with fast and light enough approaches to reproduce a large sample of the expected population, but enough refined to be reliable.
In \citet{Bandiera:2023} a first step to better include the reverberation phase in the description of the PWNe evolution has been made, based on the modification of the standard one-zone description.
Nevertheless these results, despite being much more accurate than previous models, still are far for being definitive. 
A 3D MHD modelling of reverberation is necessary to understand which role the third spatial dimension, and the structure of the magnetic field, play in shaping the PWN during its interaction with the SNR.

Moreover one-zone models by construction cannot account for the formation of asymmetric systems, that we expect will constitute the largest part of middle aged to evolved PWNe. Then, models must be somehow generalized to account for different geometries, but how without running expensive 3D -- or even 2D -- models is absolutely unclear, with only few preliminary studies presented for the moment (e.g. \citealt{Olmi_Torres:2020}). This will also be the flip of the coin for the interpretation of gamma-ray data.

\section{CONCLUSIONS}
\label{sec:conclusion}
Pulsar wind nebulae are extremely fascinating systems, showing a large variety of intriguing properties that require complex physics to be interpreted.
They are known to be powerful and efficient particle accelerators and antimatter factories in the Galaxy, maybe the primary source of the positron excess in the cosmic ray spectrum.
They are also the largest class of gamma-ray emitting sources in the Galaxy, possibly both at very high energies ($\geq 100$ GeV) and extremely high energies ($\geq 100$ TeV). 
Evolved PWNe are now known to be associated with the efficient leakage of particles in the ambient medium, showing up in two distinct ways: elongated, thin and asymmetric X-ray misaligned tails, originating from the head of bow shock PWNe; diffuse and very extended TeV halos, for the moment detected around few evolved pulsars.

Understanding and modelling the properties of PWNe in their late evolutionary phases passes through the correct modelling of all their previous  stages. Here we have reviewed what we have learned about these different phases, both from the observational and theoretical points of view, with focus on state of the art numerical modelling.
In Section \ref{sec:evo} we have in particular posed the bases for the description of the different stages of a PWN evolution, discussing from a more qualitative point of view what characterizes the four phases in which we can roughly divide it: the free-expansion phase (Sec.~\ref{sub:freeEXP}), the reverberation phase (Sec.~\ref{sub:rev}), the transitional phase after reverberation (Sec.~\ref{sub:postrev}) and the final phase outside the SNR (Sec.~\ref{sub:bspwne}).
In Section \ref{sub:other} we also reviewed what we know about PWNe in other environments, showing how these systems are prototypical of many other high-energy astrophysical sources.
A more quantitative discussion about how PWNe through their phases have been modeled can be found in Section \ref{sec:theory}. 
Here we reviewed all the different approaches used up to present days, and highlighted the main results from recent 3D MHD numerical simulations of young and old (bow shocks) PWNe (see Sec.\ref{sub:highl_young} and \ref{sub:highl_bs}).
A description of the radiation mechanisms producing the observed emission, and what we have understood about the underlying acceleration mechanisms can be found in Sec.~\ref{sec:rad_and_acc}.
PWNe observational properties, and the differences between the various phases, was discussed in Section \ref{sec:obs}.
Finally, in Section \ref{sec:open_and_where} we present our view about observational and theoretical/numerical prospects.

New impetus was already impressed in last years by the observation of TeV halos, misaligned X-ray tails and, very recently, by the detection of numerous PeVatrons in the Galaxy by LHAASO. We believe that many other new clues will come with the next generation of IACTs, as CTA or the ASTRI Mini-Array.
A very important challenge for the high energy astrophysics community will then be the interpretation of new gamma-ray data in the coming future: this might finally help solving many of the questions that remains unanswered in the fascinating pulsar wind nebulae zoo.

\begin{acknowledgements}
The authors thank Elena Amato, Rino Bandiera, Giovanni Morlino and Luca Del Zanna for the continuous collaboration, stimulating discussion and numberless coffee breaks through the years. They also acknowledge discussion with Diego F. Torres on many of the arguments treated in this review.
A special thanks goes to Rino Bandiera, who helped us with the revision of the manuscript. 
The authors also wish to acknowledge financial support from the INAF grants MAINSTREAM 2018, SKA-CTA, PRIN-INAF 2019 and from the ASI-INAF grant n.2017-14-H.O.
\end{acknowledgements}

\appendix
\section{A catalog of PWNe}
\label{sec:cat}
In this Appendix we collect the information about all the identified Galactic PWNe that we were able to find in the literature, considering  different observational bands. Here we only report those systems for which the association with a PWN seems clear, not including the large number of sources marked as possible PWNe. Instead a catalog of these systems can be found for example in \citet{Kargaltsev:2013}.

The source of the various information is specified in the caption of the two Tables.
In particular we decided to report separately PWNe clearly associated with fast moving pulsars (in Table \ref{tab:cat_old}) and all the others (in Table \ref{tab:cat_young}).
\begin{landscape}
\begin{table}
\caption{List of detected PWNe with associated PSR. Values for $P,\,\dot{P},\, B,\,\dot{E}$ and the distance $d$ are taken from the ATNF catalogue, version 1.67. For the X-ray luminosity we report, when available, the measure in the $2.1-10$ keV band (from the Chandra catalog of Galactic sources: \url{hea-www.harvard.edu/ChandraSNR/snrcat_gal.html}), otherwise the $0.5-8$ keV data from \citet{Kargaltsev:2013,Kargaltsev:2017}. In some case the luminosity in the $2.1-10$ keV band is not limited to the PWN and there is a possible contamination from the SNR (marked with a $t$ apex, standing for \textit{total}).
If the PWN has been observed in other bands, the information is given in the last column, with:  R for radio, O for optical, $\gamma$ for gamma-rays, and H$_\alpha$ (data from ``\textit{the Pulsar Wind Nebula Catalog}'' -- \url{www.physics.mcgill.ca/~pulsar/pwncat.html} and the TeVCat catalog -- \url{tevcat2.uchicago.edu}, to which we refer for updated references on the instruments detecting the various sources at gamma-rays). 
A question mark at the apex indicates a non clear association and detection or an uncertain measure.}\label{tab:cat_young}
  \fontsize{9.5}{7.8}\selectfont
  \begin{tabular}{lccccccccccc}
    \hline
    \multirow{3}{*}{\textbf{\#}} &
      \multicolumn{1}{c}{\textbf{SNR}} &
      \multicolumn{1}{c}{\textbf{PWN}} &
      \multicolumn{1}{c}{\textbf{PSR}} &
      \multicolumn{1}{c}{$\mathbf{P}$} &
      \multicolumn{1}{c}{$\mathbf{\dot{P}}$} &
      \multicolumn{1}{c}{$\mathbf{B}$} &
      \multicolumn{1}{c}{$\mathbf{\dot{E}}$} &
      \multicolumn{1}{c}{$\mathbf{d}$} &
      \multicolumn{1}{c}{$\mathbf{L_{X}}$} &
      \multicolumn{1}{c}{\textbf{band}} &
      \multicolumn{1}{c}{\textbf{also seen}}  \\
      & {  } &{  } & { } & {s} & {$10^{-14}\,\mathrm{ss}^{-1}$} & $10^{12}\,${G} & {erg s$^{-1}$} & {kpc} & {erg s$^{-1}$} & {keV} & {@}  \\
      \hline
      &  & &  &   &  &  &  &  \\
    1& G184.6-05.8 & Crab & J0534+2200 & 0.03339 & $42.10$ &  $3.79$ & $4.5 \!\times\!\! 10^{38}$ & 2 & $ 6.01\!\times\!\! 10^{36} $ & $2.1-10$ & R, O, $\gamma$\\
    2& G130.7+3.0   & 3C 58 & J0205+6449 & 0.06572 &   $19.38$ &   $3.61$ &    $2.7\!\times\!\!10^{37}$  & 3.2 &  $1.10\!\times\!\!10^{34}$ & $2.1-10$ & R, $\gamma$\\
    3& G180.0–1.7 & G179.72–1.69& J0538+2817 &   0.1432 &  $0.3669$ &  0.733  & $4.9\!\times\!\!10^{34}$ &  1.3 & $1.096\!\times\!\!10^{31}$ & $0.5-8$ & R$^?$ \\
    4& G230.4-01.4$^?$& G230.39–1.42 &  J0729-1448 &   0.2517  & $11.33 $ &  5.4   & $2.8\!\times\!\!10^{35}$ &  2.7 & $1.58\!\times\!\!10^{31}$ & $0.5-8$ & $\gamma^?$ \\
    5& G263.9–3.3 & Vela & J0835-4510 &  0.08933  & $12.50$ &  3.38 &  $6.9\!\times\!\!10^{36} $ &  0.28 &  $4.20\!\times\!\!10^{32}$  & $2.1-10$ & R, $\gamma$ \\
    6& G284.0-01.8 & G284.08–1.88 & J1016-5857 &   0.1074  & $8.083$ &  2.98  & $2.6\!\times\!\!10^{36}$ & 3.16 & $1.84\!\times\!\!10^{32}$ & $2.1-10^t$ & $\gamma^?$ \\
    7& G287.4+00.5 & Puppy & J1048-5832 &  0.1237  & $9.612$ &   3.49 &  $2\!\times\!\!10^{36}$ &  2.9 & $ 8.44\!\times\!\!10^{31} $ & $2.1-10^t$ & $\gamma$ \\
    8& G292.2-00.5 & G292.15–0.54 & J1119-6127 &  0.408 &  $402.0$ &   41 &   $2.3\!\times\!\!10^{36}$ &  8.4 & $ 2.71\!\times\!\!10^{33} $ & $2.1-10^t$ & $\gamma^?$ \\
    9& G292.0+1.8 & G292.04+1.75 & J1124-5916 &  0.1355  & $75.25 $ &  10.2 &  $1.2\!\times\!\!10^{37} $ &  5 & $ 1.36\!\times\!\!10^{35} $ & $2.1-10^t$ & R, $\gamma^?$ \\
    10& G304.1-00.2 & G304.10–0.24 & J1301-6305 &  0.1845 &   $26.67$  & 7.1 &  $1.7\!\times\!\!10^{36}$ & 10.7 & $1.45\!\times\!\!10^{32}$ & $0.5-8$ &  $\gamma$ \\
    11& G309.9-02.5 & G309.92–2.51 & J1357-6429 &  0.1661  & $36.02$ &   7.83 &  $3.1\!\times\!\!10^{36}$ &  3.1 & $8.26\!\times\!\!10^{31}$ & $2.1-10^t$ & R, $\gamma$ \\
    12& G313.6+00.3 & Kookaburra & J1420-6048 &  0.06818  & $8.3167$ &   2.41 &  $1\!\times\!\!10^{37}$ &  5.6 &  $1.4\!\times\!\!10^{33}$ & $0.5-8$ & R, $\gamma$ \\
    13& G313.3+00.1 & Rabbit & J1418-6058 &  0.11057  & $16.94$ &   4.38 &  $4.9\!\times\!\!10^{36}$ &  2-5 & $2.56\!\times\!\!10^{33}$ & $2-10^t$ & R, $\gamma$ \\
    14& G320.4–1.2 & Jellyfish & J1513-5908 &   0.1516  & $152.9$ &   15.4 &   $1.7\!\times\!\!10^{37} $ &  4.4 & $ 1.71\!\times\!\!10^{35} $ & $2.1-10$ & R, $\gamma$ \\
    15& G332.4-00.4 & G332.50–0.28 & J1617-5055 &   0.06936  & $13.51$ &   3.1  & $ 1.6\!\times\!\!10^{37}$ &  4.7 & $ 6.08\!\times\!\!10^{33} $ & $2.1-10$ & -- \\
    16& G344.7-00.1 & G344.74+0.12 & J1702-4128 &  0.1821   & $5.234 $ &  3.12  & $3.4\!\times\!\!10^{35}$ &  4.0  & $ 4.84 \!\times\!\!10^{35}  $ & $2.1-10^t$ & $\gamma^?$\\
    17& G348.9-00.4$^?$ & G348.95–0.43 & J1718-3825 &   0.07467 &  $1.322$ &   1.01 &  $1.3\!\times\!\!10^{36}$ &  3.5 & $3.98\!\times\!\!10^{32}  $ & $0.5-8$ & -\\
    18& G034.0+20.2 & G34.01+20.27 & J1740+1000 &   0.1541  & $2.147$ &   1.84  & $2.3\!\times\!\!10^{35}$ &  1.2  & $1.21\!\times\!\!10^{31}  $ & $2.1-10^t$ & R\\
    19& G008.3+00.1 & G8.40+0.15 & J1803-2137 & 0.1337 &   $13.44$ &   4.29 &  $2.2\!\times\!\!10^{36}$ &  4.4 & $2.8\!\times\!\!10^{32}  $ & $2.1-10^t$ & $\gamma^?$\\
    20& G011.1+00.1 & G11.09+0.08 & J1809-1917 & 0.08276 &   $2.553$ &   1.47  & $1.8\!\times\!\!10^{36}$ &  3.3 & $4.58\!\times\!\!10^{32}  $ & $2.1-10^t$ & R$^?$, $\gamma^?$\\
    21& G011.2-00.3 & Turtle & J1811-1925 & 0.06467 &   $4.400$  & 1.71 &  $6.4\!\times\!\!10^{36}$ &  5.0 & $8.94\!\times\!\!10^{34}  $ & $2.1-10^t$ & R, $\gamma^?$\\
    22& G018.0-00.6 & G18.00–0.69 & J1826-1334  & 0.1015 &   $7.525$  & 2.8 &  $2.8\!\times\!\!10^{36}$ &  3.6 & $4.69\!\times\!\!10^{32}  $ & $2.1-10$ & R, $\gamma$\\
    23& G21.5-0.9 & G21.50–0.89 & J1833-1034 &   0.06188  &  $20.20$ &   3.58 &   $3.4\!\times\!\!10^{37}$ &  4.1 & $2.19\!\times\!\!10^{35}  $ & $2.1-10$ & R, $\gamma$\\
    24& G029.7-00.2 & Kes75 & J1846-0258 &  0.3266  &  $710.7$ &   48.8   & $8.1\!\times\!\!10^{36}$ &  5.8 & $1.40\!\times\!\!10^{36}  $ & $2.1-10$ & R, $\gamma$\\
    25& G034.7–0.4 & G34.56-0.50 & J1856+0113 &   0.2674 &  $20.84$ &  7.55 &  $4.3\!\times\!\!10^{35}$ & 3.3 & $4.97\!\times\!\!10^{32}  $ & $2.1-10$ & R\\
    26& G054.1+0.3 & G54.10+0.27 & J1930+1852 &  0.1369  & $75.06$  & 10.3 &  $1.2\!\times\!\!10^{37}$ & 7 &$2.03\!\times\!\!10^{34}  $ & $2.1-10^t$ & R, $\gamma^?$\\
    27& G047.3-03.8 & G47.38–3.88 & J1932+1059 &  0.2265  & $0.1157$ &  0.518 &  $3.9\!\times\!\!10^{33}$ & 0.31 &$9.89\!\times\!\!10^{29}  $ & $2.1-10^t$ & R$^?$\\
    28& G075.2+00.1 & Dragonfly & J2021+3651 &   0.1037  &  $9.572$ &   3.19   & $3.4\!\times\!\!10^{36}$ &  1.8 & $7.41\!\times\!\!10^{32}  $ & $2.1-10$ & R-$\gamma^?$\\
    29& G106.6+02.9 & Boomerang & J2229+6114 &  0.05162 &   $7.827$ &  2.03 &  $2.2\!\times\!\!10^{37}$ & 3.0  &  $7.3\!\times\!\!10^{32}$ &  $2.1-10$ & R, $\gamma^?$\\
    30& G012.8-0.00  & G12.82-0.02  & J1813-1749 &  0.04474 &   $12.70$ &  2.41 &  $5.6\!\times\!\!10^{37}$ & 6.15  &  $5.35\!\times\!\! 10^{34}$ &  $2.1-10^t$ & $\gamma^?$\\
    31& G310.6-1.6  & G310.6-1.6 & J1400-6325 &  0.03118 &   $3.890$ &  1.11 &  $5.1\!\times\!\!10^{37}$ & 7.0  &  $1.13\!\times\!\!10^{35}$ &  $2.1-10^t$ & R\\
    32& SNR W42  & G25.24–0.19 & J1838-0655 &  0.0705  &   $4.925$ &  1.89 &  $5.5\!\times\!\!10^{36}$ & 6.6  &  -- &  -- & $\gamma$\\
    33& G119.5+10.2  & CTA1 & J0007+7303 &  0.3159  &   $36.00$ &  10.8 &  $4.5\!\times\!\!10^{35}$ & 1.5  &  $2.45\!\times\!\!10^{31}$ &  $2.1-10^t$ & $\gamma$\\
    34& G007.5-01.7$^?$ & Taz & J1809-2332 &  0.1468   &   $3.442$ & 2.27 &  $4.3\!\times\!\!10^{35}$ & 0.88-2$^{\mathrm{?}}$  &  $1.27\!\times\!\!10^{33}$ &  $2.1-10^t$ & R\\
    35& G266.9-01.0 & G266.97-1.00 & J0855-4644 &  0.06469   &   $0.7263$ & 0.694 &  $1.1\!\times\!\!10^{36}$ & 0.5-5.6$^{\mathrm{?}}$  &  $1.04\!\times\!\!10^{32}$ &  $2.1-10^t$ & R\\
    36& G021.9-00.1$^?$& G21.88-0.1 & J1831-0952 &  0.06727   &   $0.8324$ & 0.757 &  $1.1\!\times\!\!10^{36}$ & 3.7  &  -- &  -- & $\gamma$\\
    37& G076.9+01.0& -- & J2022+3842 &  0.04858   &   $8.610$ & 2.07 &  $3.0\!\times\!\!10^{37}$ & 7-10$^?$  &   $4.42\!\times\!\!10^{33}$  &  $2.1-10^t$ & R\\
      &  & &  &   &  &  &  &  \\    
      \hline
    \end{tabular}
\end{table}
\end{landscape}
\begin{landscape}
\begin{table}
\caption{List of the known pulsars with high proper motion and an associated PWN. Pulsar values are taken, as before, from ATNF catalogue, version 1.67, while the X-ray luminosity is always taken from \citet{Kargaltsev:2013} and \citet{Kargaltsev:2017}. Symbols and notation are the same defined in the previous table.}\label{tab:cat_old}
  \fontsize{9.5}{7.8}\selectfont
  \begin{tabular}{lccccccccccc}
    \hline
    \multirow{3}{*}{\textbf{\#}} &
      \multicolumn{1}{c}{\textbf{PWN/ASSOCIATED OBJ}} &
      \multicolumn{1}{c}{\textbf{PSR}} &
      \multicolumn{1}{c}{$\mathbf{P}$} &
      \multicolumn{1}{c}{$\mathbf{\dot{P}}$} &
      \multicolumn{1}{c}{$\mathbf{B}$} &
      \multicolumn{1}{c}{$\mathbf{\dot{E}}$} &
      \multicolumn{1}{c}{$\mathbf{d}$} &
      \multicolumn{1}{c}{$\mathbf{L_{X}}$} &
      \multicolumn{1}{c}{\textbf{band}} &
      \multicolumn{1}{c}{\textbf{also seen}}  \\
      & {  } &{  } & {s} & {$10^{-14}\,\mathrm{ss}^{-1}$} & $10^{12}\,${G} & {erg s$^{-1}$} & {kpc} & {erg s$^{-1}$} & {keV} & {@}  \\
      \hline
      &  &   &    &   &    &  &  & &  &  \\
     1 & Geminga &  J0633+1746 &   0.2371 &  $1.097$ &   1.63 &  $3.2\!\times\!\!10^{34}$ & 0.19 & $2.24\!\times\!\!10^{29}$ & $0.5-8$ & $\gamma$ \\
     2 & G319.97-0.62 & J1509-5850 &   0.08892 &  $0.9166$ &   0.914 &   $5.1\!\times\!\!10^{35}$ &  3.4 &  $1.12\!\times\!\!10^{33}$ & $0.5-8$ & R \\
     3 & G343.10-2.69 & J1709-4429 & 0.1025  & $9.298$ &  3.12 & $3.4\!\times\!\!10^{36}$ &  2.6&  $3.98\!\times\!\!10^{32}$ & $0.5-8$ & R, $\gamma^?$ \\
     4 & Mouse & J1747-2958 & 0.09881 &  $6.132$ &   2.49  & $2.5\!\times\!\!10^{36}$ &  2.5 &$6.76\!\times\!\!10^{33}$ & $0.5-8$ & R \\
     5 & Duck & J1801-2451 & 0.1249 &  $12.79$ &   4.04  & $2.6\!\times\!\!10^{36}$ & 3.8 & $1.6\!\times\!\!10^{33}$ & $0.5-8$ & R, $\gamma$ \\
     6 & Guitar & J2225+6535 &  0.6825 &  $0.9661$ &   2.6 &  $1.2\!\times\!\!10^{33}$ &  0.8$^a$ & $1.51\!\times\!\!10^{30}$ & $0.5-8$ & H$_\alpha$ \\
     7 & Morla & J0357+3205 &   0.4441  & $1.304 $ &   $2.43 $ &  $5.9\!\times\!\!10^{33}$ &  0.8 & $1.17\!\times\!\!10^{30}$ & $0.5-8$ & -- \\
     8 & -- & J0437-4715 &   0.005757  & $5.7\!\times\!\!10^{-6} $ &   $5.81\!\times\!\!10^{-4} $ &  $1.2\!\times\!\!10^{34}$ &  0.16 & $\sim 4\!\times\!\!10^{28}$ & $0.5-8$ & H$_\alpha$, R$^{\mathrm{?}}$ \\
     9 & SNR S147 & J0538+2817 &   0.1432   & $0.3669 $ &   $0.733 $ &  $4.9\!\times\!\!10^{34}$ &  1.3 & $2\!\times\!\!10^{31}$ & $0.5-8$ & R \\     
     10 & -- & J0742-2822 &   0.1668   & $1.682 $ &   $1.69 $ &  $1.4\!\times\!\!10^{35}$ &  2 & -- & -- & H$_\alpha$, R$^{\mathrm{?}}$ \\ 
     11 & -- & J0908-4913 &   0.1068   & $1.51 $ &   $1.28 $ &  $4.9\!\times\!\!10^{35}$ &  1 & -- & -- & H$_\alpha^{\mathrm{?}}$, R \\ 
    12 & Lighthouse & J1101-6101 &   0.0628  & $0.86 $ &   $0.742$ &  $1.4\!\times\!\!10^{36}$ &  7 & $2.5\!\times\!\!10^{32}$ & $0.5-8$ & H$_\alpha^{\mathrm{?}}$ \\ 
    13 & G293.79+0.58 & J1135-6055 &   0.1149  & $7.93 $ &   $3.05 $ &  $2.1\!\times\!\!10^{36}$ &  $2.9$ & $2.5\!\times\!\!10^{32}$ & $0.5-8$ & H$_\alpha^{\mathrm{?}}$, R$^{\mathrm{?}}$ \\ 
    14& Frying Pan & J1437-5959 &   0.0617  & $0.8587$ &   $0.737 $ &  $1.4\!\times\!\!10^{36}$ &  8.5 & -- & -- & R \\ 
    15 & G006.4+04.9 & J1741-2054 &   0.4137   & $1.698$ &   $2.68 $ &  $9.5\!\times\!\!10^{33}$ &  0.3 & $1.6\!\times\!\!10^{30}$ & $0.5-8$ & -- \\ 
    16 & Eel & J1826-1256 &   0.1102   & $12.15$ &   $3.7 $ &  $3.6\!\times\!\!10^{36}$ &  1.6 & $2.4\!\times\!\!10^{33}$ & $0.5-8$ & H$_\alpha^{\mathrm{?}}$, $\gamma^?$ \\   
    17 & SNR W44 & J1856+0113 &   0.2674   & $20.84$ &   $7.55 $ &  $4.3\!\times\!\!10^{35}$ &  3.3 & $1.6\!\times\!\!10^{33}$ & $0.5-8$ & R \\    
   18 & G47.38-3.88 & J1932+1059 &   0.2265   & $0.1157$ &   $0.518 $ &  $3.9\!\times\!\!10^{33}$ &  0.3 & $3.16\!\times\!\!10^{29}$ & $0.5-8$ & H$_\alpha^{\mathrm{?}}$, R \\ 
    19 & SNR CTB 80 & J1952+3252 &   0.03953   & $0.5845$ &   $0.486 $ &  $3.7\!\times\!\!10^{36}$ &  3 & $1\!\times\!\!10^{33}$ & $0.5-8$ & H$_\alpha$, R \\   %
    20 & Black Widow & J1959+2048 &   0.001607   & $1.69\!\times\!\!10^{-6} $ &   $1.67\!\times\!\!10^{-4} $ &  $1.6\!\times\!\!10^{35}$ &  1.4 & $5.4\!\times\!\!10^{29}$ & $0.5-8$ &  R\\  
    21 & -- & J2030+4415 &   0.2271   & $0.6484$ &   $1.23 $ &  $2.2\!\times\!\!10^{34}$ &  0.7 & $3.1\!\times\!\!10^{30}$ & $0.5-8$ & H$_\alpha$\\
    22 & -- & J2055+2539 &   0.3196   & $0.408$ &   $1.16 $ &  $4.9\!\times\!\!10^{33}$ &  0.6 & $1.48\!\times\!\!10^{30}$ & $0.5-8$ & H$_\alpha^{\mathrm{?}}$, R$^{\mathrm{?}}$\\   
    23 & G10.92-45.43 & J2124-3358 &   0.004931   & $2.06\!\times\!\!10^{-6} $ &   $3.22\!\times\!\!10^{-4} $ &  $6.8\!\times\!\!10^{33}$ &  0.4 & $9.5\!\times\!\!10^{28}$ & $0.5-8$ & R\\ 
    24 & Mushroom & J0358+5413 &   0.1564 & $0.4395$ &  0.839 &  $4.5\!\times\!\!10^{34}$ & 1 &  $1.58\!\times\!\!10^{31}$ & $0.5-8$ & --\\
    25 & -- & J1648–4611 &   0.165  & $2.373$ &  2.0 &  $2.1\!\times\!\!10^{35}$ & 4.5 &  $<3\!\times\!\!10^{31}$ & $0.5-8$ & --\\
     &  &   &    &   &    &  &  & &  &  \\
    \hline
     \multirow{2}{*} 
      & \\
     & \footnotesize{$^a\,$Updated distance from \citet{Deller:2019}.} \\ 
  \end{tabular}
\end{table}
\end{landscape}
\bibliographystyle{pasa-mnras}
\bibliography{biblio}

\end{document}